\documentclass[
  journal=pasa,
  manuscript=article-type,
  year=202X,
  volume=XX,
]{cup-journal}

\usepackage{amsmath}
\usepackage{amssymb,microtype,siunitx,booktabs}
\usepackage{graphicx}        

\usepackage{color} 
\usepackage{subfig}
\usepackage{float}
\usepackage{stfloats}

\usepackage{hyperref}
\hypersetup{
    colorlinks=true,
    linkcolor=blue,
    filecolor=magenta,      
    urlcolor=blue,
    citecolor=blue
    }

\chardef\us=`\_

\title{STORMY : A Real-time Triggering Framework using Yamagawa Solar Spectrograph for Active Solar Emission Observations with the MWA}

\author{Deepan Patra}
\affiliation{National Centre for Radio Astrophysics, Tata Institute of Fundamental Research, S. P. Pune University Campus, Pune, India, 411007}
\author{Devojyoti Kansabanik}
\affiliation{NASA Jack Eddy Fellow, University Corporation for Atmospheric Research, 3090 Center Green Dr., Boulder, CO, USA, 80301}
\alsoaffiliation{Johns Hopkins University Applied Physics Laboratory, 11001 Johns Hopkins Rd, Laurel, MD, USA 20723}
\author{Divya Oberoi}
\affiliation{National Centre for Radio Astrophysics, Tata Institute of Fundamental Research, S. P. Pune University Campus, Pune, India, 411007}
\author{Yûki Kubo}
\affiliation{National Institute of Information and Communications Technology, Tokyo, Japan}
\author{Bradley W. Meyers}
\affiliation{Australian SKA Regional Centre (AusSRC), Curtin University, Bentley, WA 6102, Australia}
\alsoaffiliation{International Centre for Radio Astronomy Research, Curtin University, Bentley, WA, Australia}
\author{Andrew Williams}
\affiliation{International Centre for Radio Astronomy Research, Curtin University, Bentley, WA, Australia}
\author{Soham Dey}
\affiliation{National Centre for Radio Astrophysics, Tata Institute of Fundamental Research, S. P. Pune University Campus, Pune, India, 411007}
\author{Naoto Nishizuka}
\affiliation{National Institute of Information and Communications Technology, Tokyo, Japan}

\begin{document}
\begin{abstract}

Some of the most interesting insights into solar physics and space weather come from studying radio emissions associated with solar activity, which remain inherently unpredictable. Hence, a real-time triggering system is needed for solar observations with the versatile new-generation radio telescopes to efficiently capture these episodes of solar activity with the precious and limited solar observing time. We have developed such a system,  \textbf{S}olar \textbf{T}riggered \textbf{O}bservations of \textbf{R}adio bursts using \textbf{M}WA and \textbf{Y}amagawa (STORMY) for the Murchison Widefield Array (MWA), the precursor for the low frequency telescope of upcoming Square Kilometre Array Observatory (SKAO).
It is based on near-real-time data from the Yamagawa solar spectrograph, located at a similar longitude to the MWA. We have devised, implemented, and tested algorithms to perform an effective denoising of the data to identify signatures of solar activity in the Yamagawa data in near real-time. End-to-end tests of triggered observations have been successfully carried out at the MWA. STORMY is operational at the MWA for the routine solar observations, a timely development in the view of the ongoing solar maximum. We present this new observing framework and discuss how it can enable efficient capturing of event-rich solar data with existing instruments, like the LOw Frequency ARray (LOFAR), Owens Valley Radio Observatory - Long Wavelength Array (OVRO-LWA) etc., and pave the way for triggered observing with the SKAO, especially the SKA-Low.
\end{abstract}


\section{Introduction}\label{S-Introduction} 
The observed solar radio emission is a superposition of emissions from a large variety of mechanisms -- from incoherent mechanisms like thermal bremsstrahlung and gyromagnetic emission \citep[e.g.,][]{white1997radio,nindos2020incoherent} to coherent emission mechanisms like plasma emission \citep[e.g.][]{ginzburg1959mechanisms,melrose1980emission} and electron-cyclotron maser emission \citep[e.g.][]{winglee1986electron,tang2013electron}. The coherent emission mechanisms are capable of producing much higher brightness temperatures and hence give rise to a large increase in the observed intensity. These bright emissions, known as solar radio bursts, are linked with various extreme solar phenomena ranging from energetic CMEs to flares spanning a large range of energies. The radio bursts originate from plasma emission and their spectro-temporal variation in dynamic spectra 
can show markedly different structures due to different origins of the energetic electrons producing these bursts. Based on their morphology in the dynamic spectrum, they are classified into several types \citep{wild,mclean1985solar}. 
Detecting and studying such solar bursts is of utmost importance for understanding their emission mechanisms \citep{bastian1998radio,melrose2017coherent}, probing the turbulent magnetized coronal medium and studying the space-weather \citep{Zhang2014}.

Owing to their high brightness, many fine details of the emission can be captured by comparatively simpler and inexpensive single dish/element instrumentation. Technological ease and the comparative simplicity of analysis have implied that non-imaging spectrograph observations have dominated the studies of the Sun by a very wide margin. Despite their obvious limitations, we have learnt a lot from spectroscopic observations \citep[e.g.,][]{maxwell_swarup,classen2002association,mann_formation,gopalswamy2005type}. However, due to the large and potentially simultaneous variations in temporal and spectral structures, brightness temperatures and polarization properties of the solar radio bursts, spectro-polarimetric snapshot imaging with high dynamic range is needed to study these events in their full glory. 
Generations of former and presently operational radioheliographs like Culgoora \citep{Wild1967}, Clark Lake \citep{Gregley1985},
NoRH \citep[][]{nakajima,takano2007upgrade,shibasaki2013long}, the Nançay Radioheliograph \citep[][]{kerdraon, saint2012decade} and Gauribidnur \citep{Ramesh1998} etc. enabled routine imaging of the Sun at a few discrete frequencies, helping in the study of solar flares and noise storms.
Despite this, the instrumentation, algorithmic and computational challenges posed by the needs of solar radio imaging have held back progress. 
More recent solar-dedicated instruments such as the Siberan Radioheliograph \citep{srh_2020}, Mingantu Spectral Redioheliograph \citep{yan2021mingantu}, Expanded Owens Valley Solar Array \citep[EOVSA;][]{gary2018microwave}, Owens Valley Radio Observatory - Long Wavelength Array \citep[OVRO-LWA;][]{chhabra2021imaging}, and Daocheng Solar Radio Telescope \citep[DSRT;][]{Yan2023-DSRT}, have made progress in addressing some of the limitations of earlier instruments  
\citep[see review by][]{gary2023new}.
Alongside, over the last decade or so, a new generation of radio interferometers have emerged, e.g. the LOFAR \citep{lofar_paper}, the MWA \citep{tingay2013murchison}, MeerKAT \citep{meerkat_2016}. 
In terms of their raw capabilities, these instruments tend to be best-in-class along some/many of the parameters of interest (e.g. uv-coverage, resolution, spectroscopic imaging capability, calibration fidelity, etc.) and are much better suited to the demanding requirements of solar imaging. These sophisticated instruments, all precursors or pathfinders for the Square Kilometre Array Observatory (SKAO), are designed to be versatile, capable of addressing a vast variety of science areas and optimized for synthesis observations of faint sources. Considerable effort is required and has been put in to adapt these instruments and the analysis procedures for solar observations \citep[e.g., ][]{Kansabanik2022c,Kansabanik_2022,Kansabanik_2023, dey2025automated,kansabanik2025solar}. These efforts have and continue to deliver considerable science returns \citep[e.g.,][and many more]{morosan_2014,reid_kontar,kansabanik2024spectropolarimetric,oberoi2023preparing,shilpi_type2,kansabanik2025solar,dey2025first}.

Being versatile and state-of-the-art, the observing time at these new generation facilities is highly oversubsribed.
On the other hand, the 
solar radio bursts are infrequent, transient, and most importantly, unpredictable. 
Hence, blind scheduled solar observations are not only an extremely inefficient way to capture these bursts in large numbers, they also lead to a high likelihood of missing big solar events.

One way to overcome this observational limitation is by optimally utilizing the available observing time using a robust and reliable automated near-real-time triggered observation framework. By enabling one to use precious observing time only when some solar activity is known to have just taken place, such a system can vastly increase the efficiency of the limited available observing time to capture instances of solar activity. Having such systems will not only help in boosting efficiency of Solar research with the existing instruments but also will pave a way to prepare for optimized solar observations with the upcoming SKAO.

In radio astronomy, triggered observations have been widely used to capture astrophysical transient activities such as FRBs, GRB afterglows, or pulsar glitches where the response times of seconds to hours are acceptable \citep[e.g.,][]{frb_trigger_2020,grb_triggered_2024}. Solar radio burst triggers need near-real time response and low-latency setups to configure the interferometer to observe the Sun as soon as the trigger is received.
We have developed such a framework for the low-frequency precursor of the SKAO, the MWA, based on near-real-time data from the Yamagawa solar radio spectrograph \citep{Iwai_2017}. We have named this framework \textbf{S}olar \textbf{T}riggered \textbf{O}bservations of \textbf{R}adio bursts using \textbf{M}WA and \textbf{Y}amagawa (STORMY).

The paper is organized as follows. An overview of the pipeline is given in section \ref{sec:overview}. Section \ref{sec:preprocess} and \ref{sec:event_detection} describe the preprocessing and the event detection, respectively. Trigger generation and observation are shown in section \ref{sec:mwa_triggering}. We present the performance test of STORMY in section \ref{sec:performance}. Section \ref{sec:final_results} shows a few events captured by our pipeline. Finally, we discuss the application of this algorithm to other facilities in section \ref{sec:application}.

\section{Overview of Real-time Burst Detection and Triggering Pipeline}\label{sec:overview}
While there have been some attempts to develop automated algorithms for the detection of solar radio bursts from dynamic spectrum, they tend to be focused on the detection and characterization of the type of bursts (e.g. type IIs or IIIs). Earlier works by \citet{Lobzinn} and \citet{bonnin2011automated} proposed the use of the Hough transform \citep{hough1962} on the dynamic spectra. \citet{Dayal} have developed an automated burst detection pipeline using e-CALLISTO \citep{benz2005callisto,benz2009world} data, and \citet{salmane} have developed a Constant False Alarm Rate (CFAR) based technique for automatic detection of radio bursts in data from the Nan\c{c}ay Decametric Array \citep{boischot1980new}. In recent times, scientists have also successfully implemented different convolutional neural network (CNN) based deep learning models to detect and categorize solar radio bursts \citep[e.g.,][]{scully2021type,orue2023automatic,deng2024real}.
However, these algorithms were developed for searching and classifying radio bursts present in the archival data, where the entire morphological information of the burst is available in the dynamic spectrum. 

In case of generating a trigger for scheduling an interferometric observation, one does not have the luxury of allowing the full solar radio event to play out before raising the trigger. Developing such a framework to conduct robust and efficient triggered observations requires the following capabilities: (1) near-real-time access to appropriate data from solar-dedicated telescopes, based on which a trigger can be generated; (2) algorithms and software tools needed to ingest and process these data, including reliably rejecting any radio frequency interference (RFI), to provide a trigger; (3) the framework required for this trigger to be communicated to the telescope under consideration and, (4) scheduling the observations and the acquisition of data. Here we focus on the first two of the aspects mentioned here, and the latter two have already been implemented at the MWA for other astronomical observations and have been adapted for the present purpose. 

\subsection{Yamagawa and MWA}\label{sec:instruments}
Our solar observing trigger is derived from the data received from the Yamagawa heliospectrograph, located in Japan at a latitude and longitude of 31.204$^{\circ}$N and 130.617$^{\circ}$E, respectively. It observes the Sun daily from rise to set in the frequency band from 70 MHz to 9 GHz. Since Yamagawa is at a similar longitude as the MWA (26.703$^{\circ}$S, 116.671$^{\circ}$E) and the MWA only observes the Sun above 30$^{\circ}$ elevation, Yamagawa has complete overlap with the solar observing window of the MWA.
\begin{figure}[!h]
    \centering
    \includegraphics[clip,width=\linewidth]{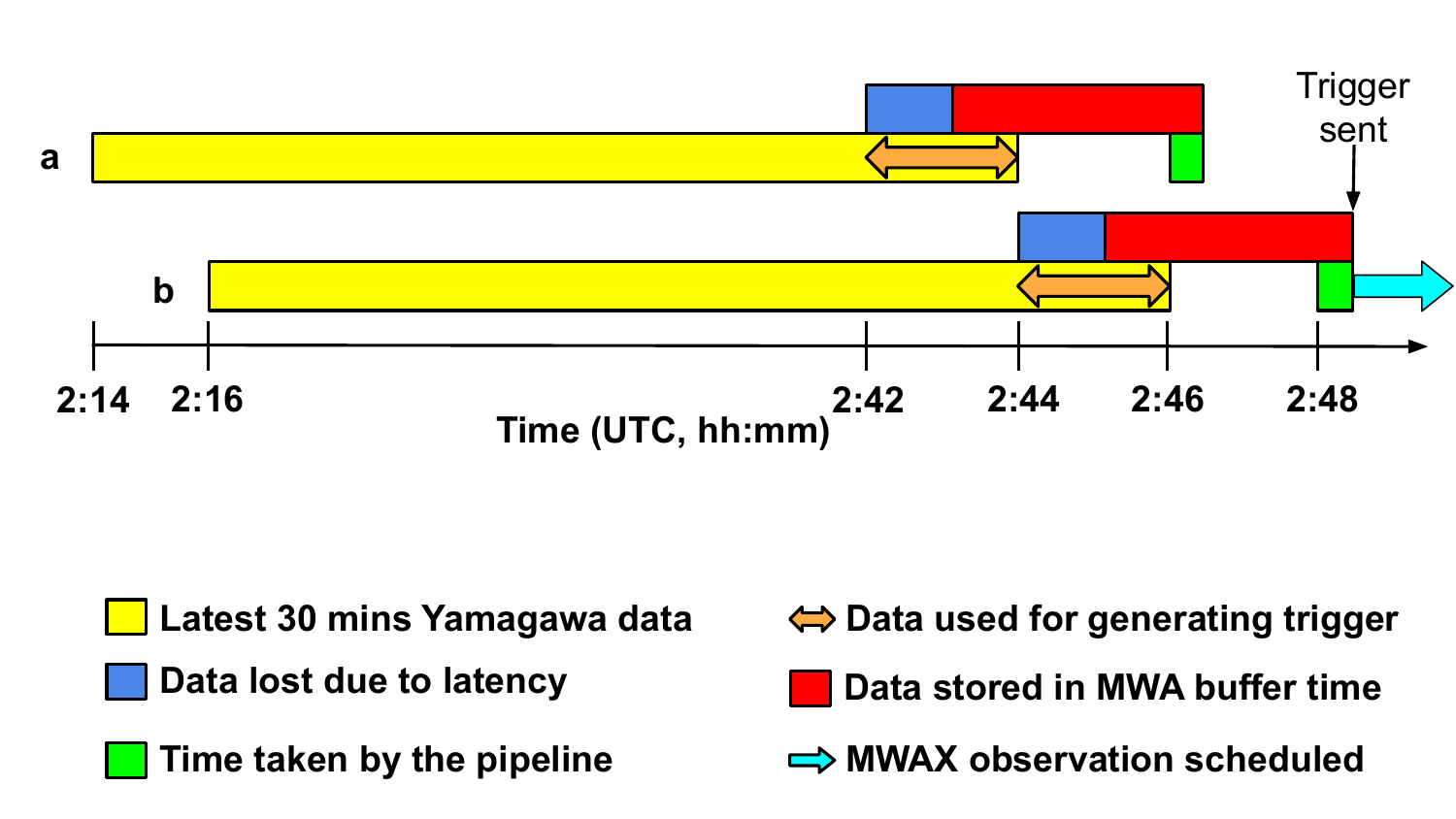}
    \caption{This figure shows an example of the real-time update of data from Yamagawa spectrograph and the MWA voltage buffer. For example: the chunk a is available at 2:46 and the next chunk b is available at 2:48. When a trigger is sent, the buffer data (red block) is saved and a regular correlator observation (blue arrow) is triggered. We lose about 70--80 s of data even after recovering 160 s of data from the MWA buffer.}
    \label{fig:data_chunk}
\end{figure}
We have optimized the Yamagawa system to enable a custom near-real-time datastream. This datastream provides dynamic spectrum at 2-minute intervals with a latency of 2 minutes and is better suited to our needs than the publicly available data. 
The data have a spectral and temporal resolution of 1 MHz and 1 second, respectively. 
\begin{figure*}[!b]
    \centering
    \includegraphics[width=0.75\linewidth]{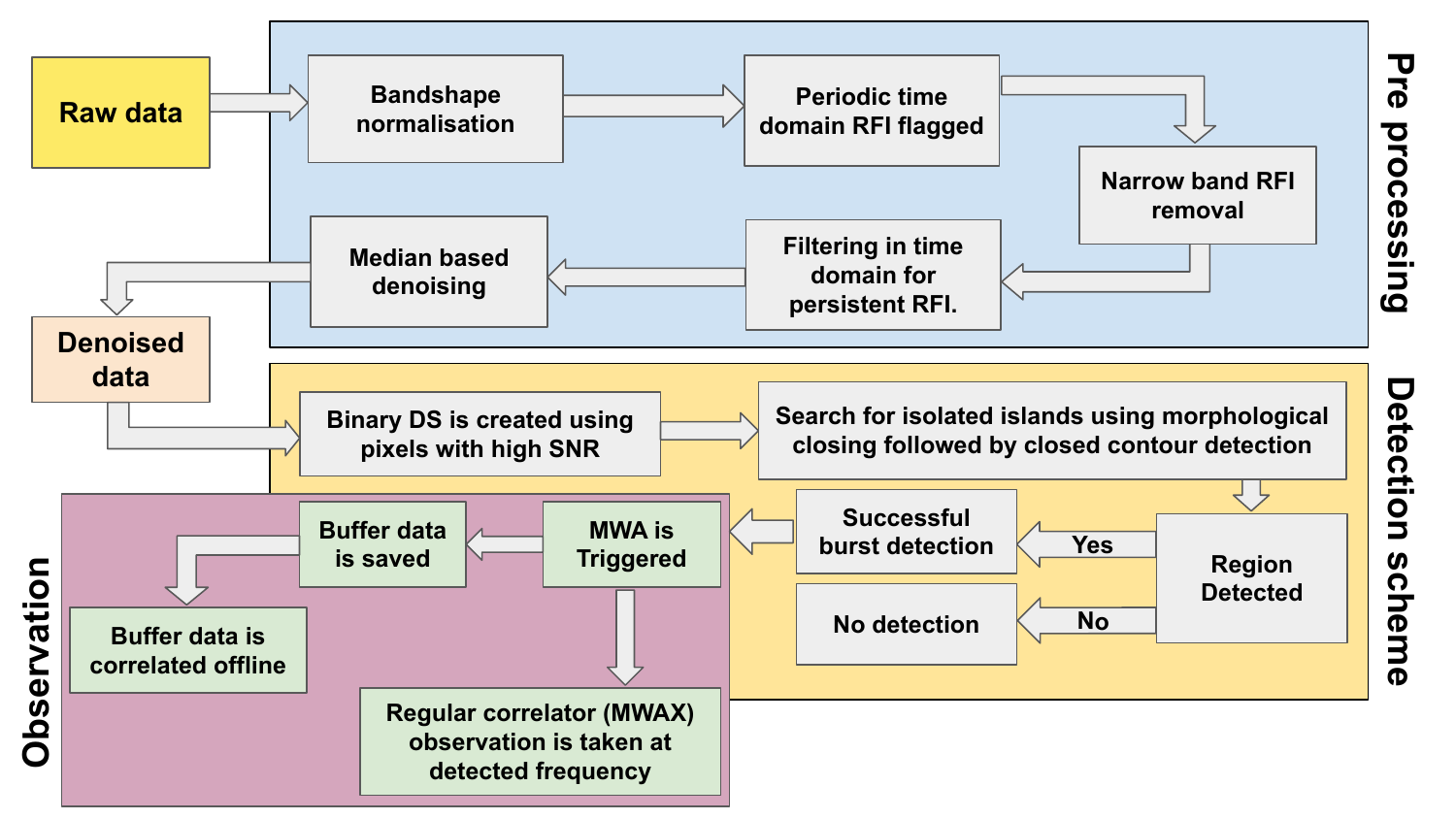}
    \caption{A schematic block diagram providing an overview of the entire observation triggering pipeline.}
    \label{fig:flowchart}
\end{figure*}

The MWA observes in the 80--300 MHz band and offers 30.72 MHz of instantaneous processed bandwidth. 
This instantaneous bandwidth is available in 24 chunks of 1.28 MHz each, which can be distributed flexibly across the entire observing frequency range. 
Solar observations are often carried in the so called {\em picket fence} mode where the observing bandwidth is distributed in 1.28 MHz chunks across the full 80--300 MHz band in a log-spaced manner.

The MWA also offers a unique ability to store the voltage data for each of the polarizations from each of the tiles in a ring buffer \citep{morrison2023mwax}.
These data are extremely voluminous ($\sim$56 TB/hour for 128 tiles) and with 128 tiles the ring buffer can store upto $160\ \mathrm s$ of raw voltage data.
The voltage buffer offers two key advantages -- first, it provides access to the data for up to $160\ \mathrm s$ prior to the trigger arriving at the MWA and thus allowing us to recover much of the event data lost due to the latency in raising the trigger; and second, it provides flexibility in the time and frequency resolution with which to compute correlations (visibilities) down to the Nyquist sampling rate.
On command, this buffer data can be dumped to disk and can later be passed through an offline correlator to obtain visibility data and reduce data volume. The buffer is filled by the datastream from the pointing of the last observation. Hence, we configure the MWA to point at the Sun in the picket fence mode for all of the unallocated daytime hours, when the Sun is above the elevation limit of the MWA (30$^{\circ}$). So even though no visibility data is recorded, this ensures that the buffer is populated with solar data.
Figure \ref{fig:data_chunk} shows how the near real-time datastream comes from the Yamagawa spectrograph, the time taken to raise the trigger and the start of observations at the MWA. It is clear from the diagram that due to the buffer available at the MWA (red block), we lose a very small amount of data (blue block) due to the latency of the Yamagawa data.

\subsection{Structure of the Pipeline}
For an overview of the full pipeline, see figure \ref{fig:flowchart}. After the near-real-time data is downloaded, the data flows through the following parts:  
\begin{enumerate}
    \item \textbf{Pre-processing :} We have implemented efficient RFI mitigation algorithms along with denoising to improve the contrast of the solar burst signals in the dynamic spectrum.
    \item \textbf{Event detection :} In this block, we find trigger-worthy feature in binary dynamic spectra obtained from the previous block.
    \item \textbf{Observation :} If any burst is detected, a trigger is sent to the MWA to schedule solar observations with appropriate observing settings.
\end{enumerate}

\section{Pre-processing of the Raw-Data}\label{sec:preprocess}
The real-time data received from the Yamagawa spectrograph has different types of RFI, e.g., narrow band but persistent, broadband but short-lived, etc. There are also periodic signals, as well as others that tend to appear only at certain times of the day. Apart from these, the raw data is also not corrected for the instrumental bandshape. 

Before detecting any solar signal, we need to remove the RFI and also correct for the intrinsic bandshape of the instrument. 

\begin{figure}[!htbp]
    \centering
    \includegraphics[width=\linewidth]{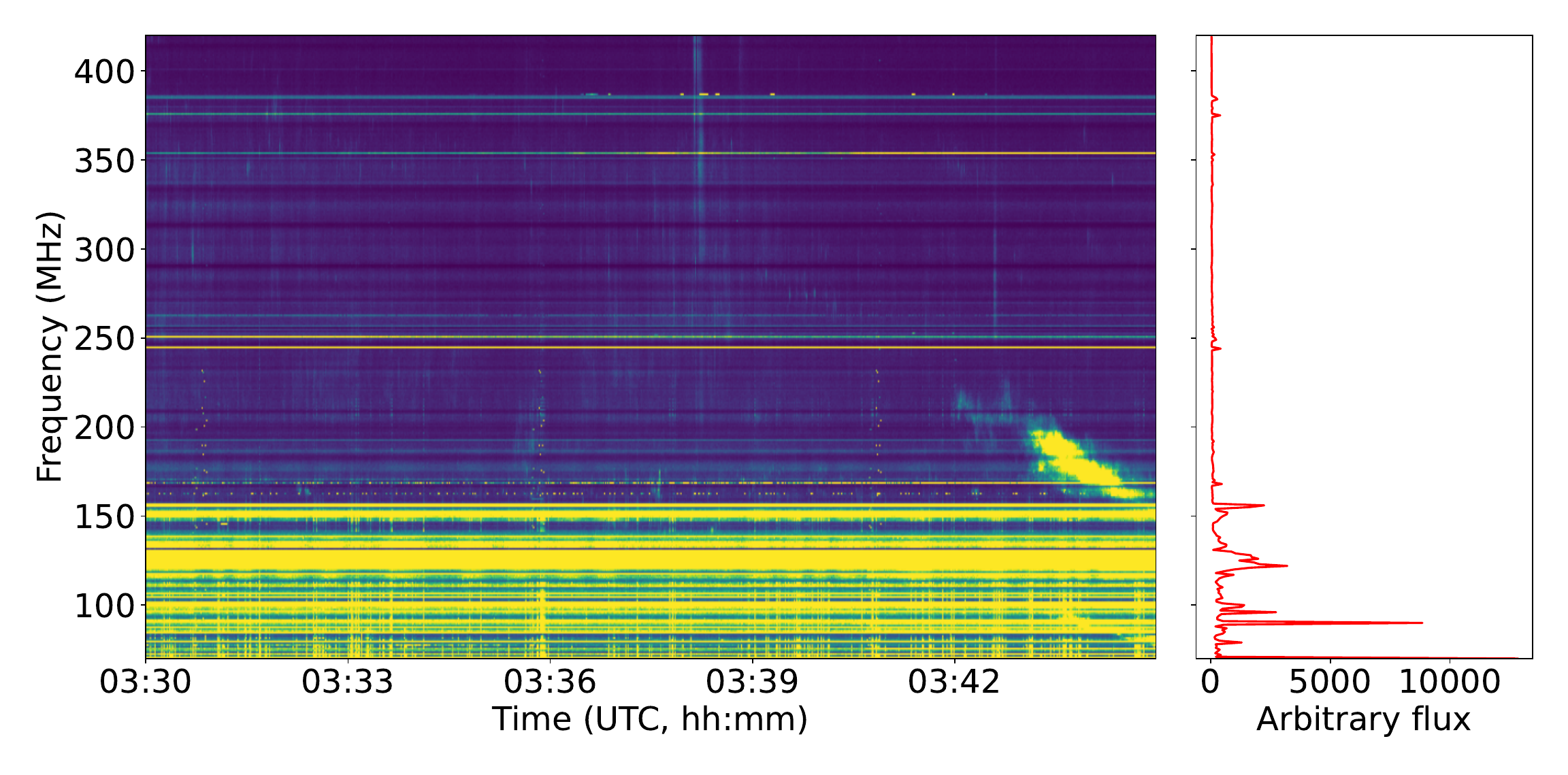}
    \caption{Raw Yamagawa data is shown on the left panel and the median bandpass calculated from previous full day data is shown in the right panel.}
    \label{fig:raw}
\end{figure}

\subsection{Bandpass Correction}\label{subsec:bandpass_cor}
The bandshape of an instrument is its spectral response and can vary significantly across the observing band. The observed intensity of the signal that is recorded by each spectral channel gets modified by the instrumental bandshape. Hence, we need to correct for the power levels of each spectral channel by the instrumental bandshape. Although the same-day data can also be used for band shape calculation, identifying quiet time without any solar bursts in the real-time data becomes computationally inefficient, and also, if a very long duration solar signal is present in the data, the median might get affected. Hence, to calculate the bandshape of the Yamagawa spectrograph, we use previous full-day data ($\sim$8 hours). This data is split into chunks each spanning $180\ \mathrm s$. Now, we calculate the bandshape of each chunk ($B_i(\nu)$) by taking median along the temporal axis and then we calculate the final bandshape by taking median of all chunks ($B(\nu)=Median[B_i(\nu)]$). This median bandshape is used to normalize the dynamic spectra as follows, $S(\nu,t)=\frac{I(\nu,t)}{B(\nu)}$, where, $I(\nu,t)$ denotes the raw dynamic spectrum. This normalization removes instrumental spectral response and solar features become more easily detectable. Figure \ref{fig:raw} shows a sample raw data and the previous day bandshape used for bandpass correction. We will use this data throughout this paper to show further steps.
\begin{figure}
    \centering
    \includegraphics[width=\linewidth]{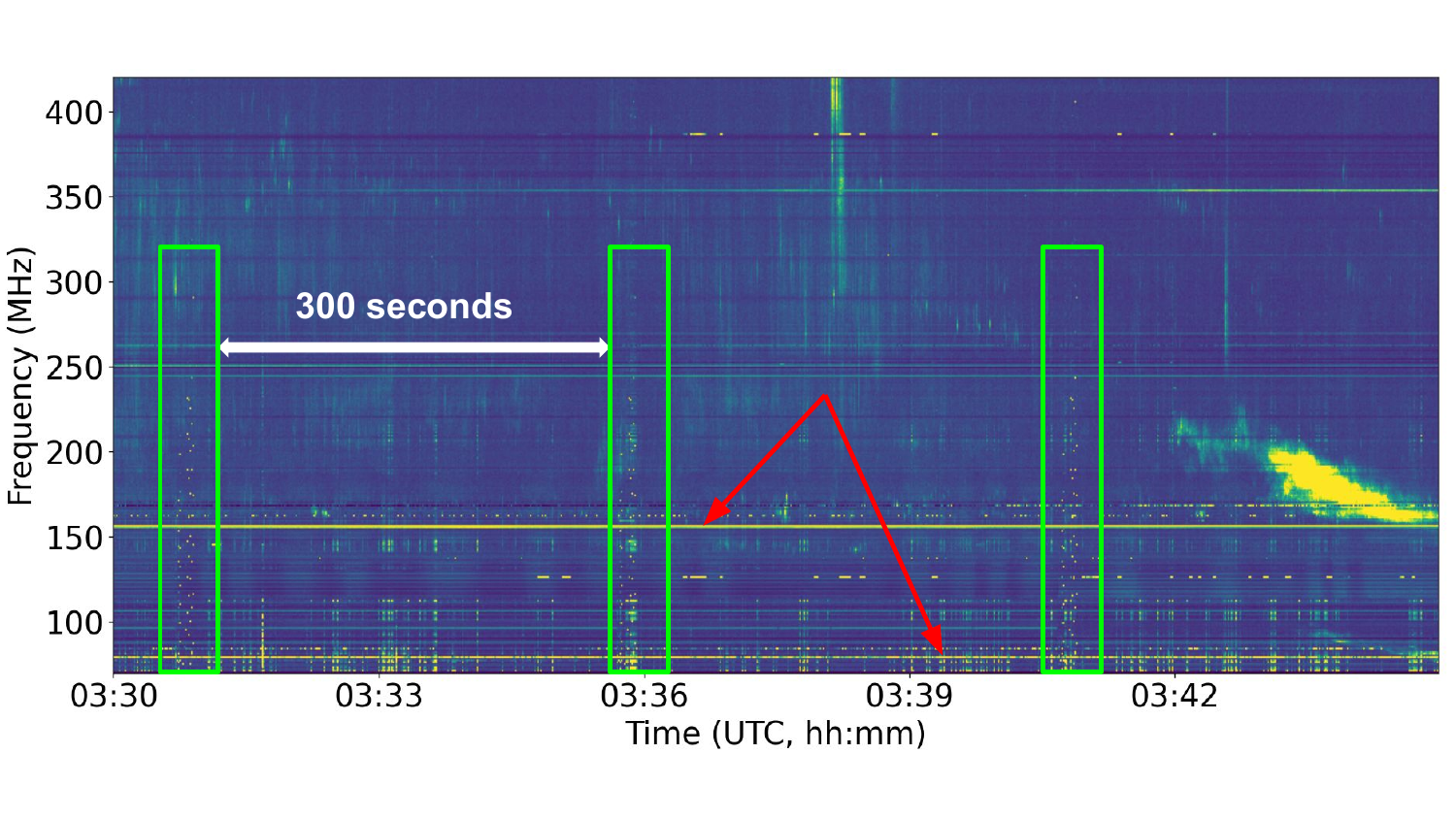}
    \caption{This figure shows the bandpass normalized dynamic spectrum where the varying signals are easily detectable. The yellow rectangles show a periodic RFI that is seen every 300 s. The red arrows show narrow band persistent RFIs.}
    \label{fig:bandpass_normalised}
\end{figure}

\subsection{RFI Mitigation}\label{subsec:rfi_mitigation}
After the bandpass normalization, different spectrally and temporally varying features can be seen with enhanced contrast. However, the dynamic spectrum has different types of unwanted contamination from RFI which can sometimes mimic the true solar signal. We need to remove them before starting to detect the solar signal. Otherwise, we may get false positive detections. Analyzing the archival Yamagawa data, we identified a few different types of RFI that can be present in the data (Figure \ref{fig:bandpass_normalised}) -- (1) periodic RFI, (2) narrow band and persistent RFI and (3) transient RFI.

\begin{figure*}[!thb]
    \centering
    \includegraphics[width=0.9\linewidth]{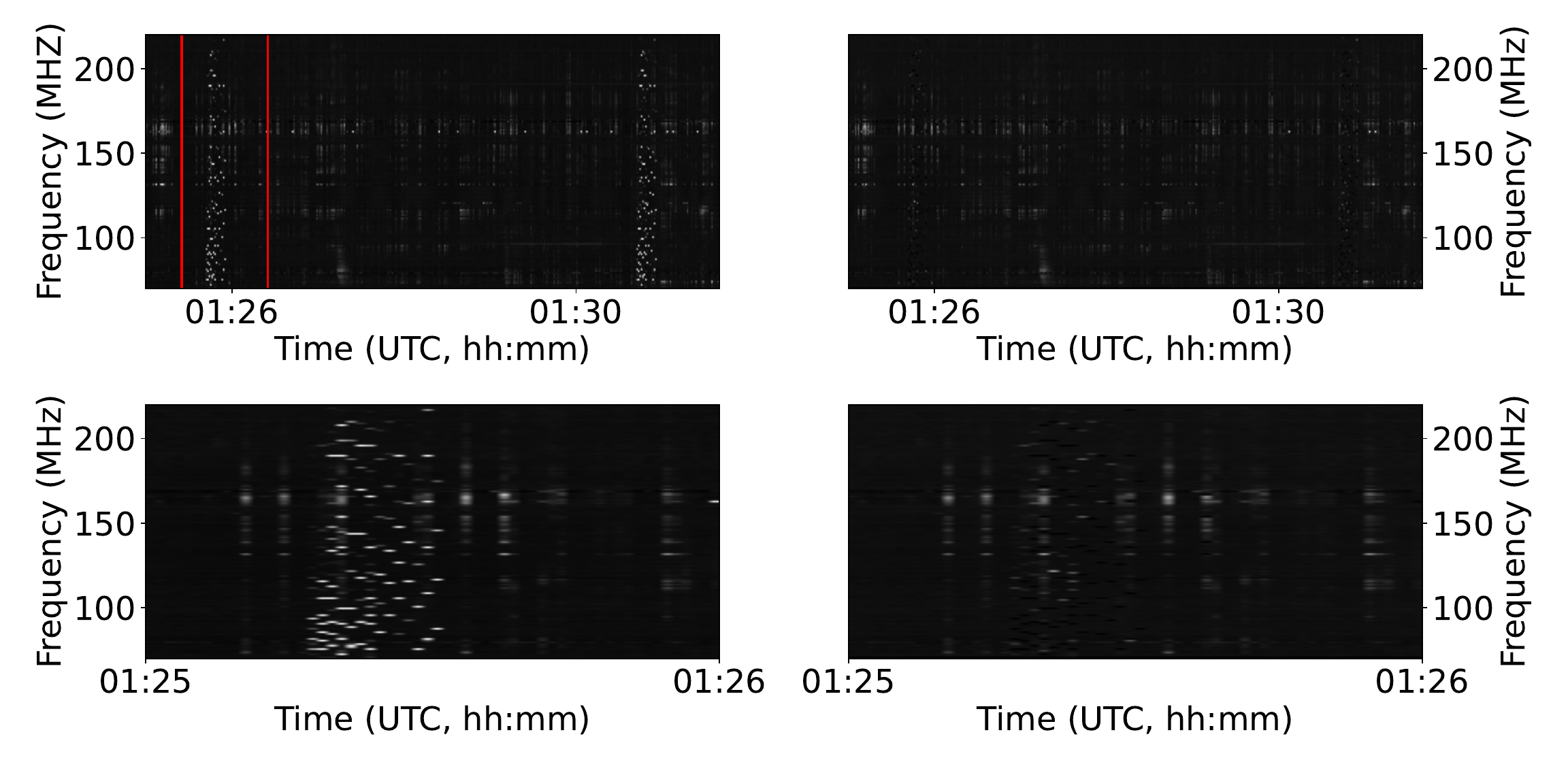}
    \caption{ This figure shows how the periodic RFI is cleaned. The top left panel shows the dynamic spectrum with fragmented periodic RFI. The top right panel shows the dynamic spectrum after cleaning these RFI. Bottom panels: Zoomed-in view of the interval between the red lines, showing before (left) and after (right) cleaning the RFI.}
    \label{fig:periodic}
\end{figure*}

First, we remove the periodic RFI. They are present every day and have a periodicity of five minutes with a duty-cycle of $\sim10\ \mathrm s$, but are fragmented across the channels (Refer to the bottom left panel of figure \ref{fig:periodic}). Although they are fragmented, they appear exactly after 300 s in each channel (top left panel of figure \ref{fig:periodic}). 
Let us denote $S_{\nu}(t)$ as time series a channel. We use the \textit{find\_peaks} module from \textit{scipy} to find peaks with periodicity of 300 s by using the following parameters:
\begin{itemize}
    \item \textbf{Distance :} Setting this to 300 to find the periodic peaks.
    \item \textbf{Width :} Setting this to 1 since the RFI is fragmented in time and frequency and the pixels are isolated.
    \item \textbf{Height :} We calculate the minimum height of peaks as:
    \begin{equation}
        h=Median(S_\nu(t))+k\sigma
    \end{equation}
\end{itemize}
here we use k=[100,50,25] to account for different levels of bright RFI. Figure \ref{fig:periodic} shows the before and after cleaning of the periodic RFI.

Next, we clean any persistent narrow-band RFI. They span only a few channels in frequency, deviating largely from the background and persistently present for a long time. Since solar emissions can also have significant deviations from the background, extra care should be taken while filtering these RFI. We used a median filter with a sufficiently small moving window (5 MHz) to arrive at a smoothed representation of the spectrum at a timestamp, $t_0$. This is denoted by $S_\mathrm{m}(\nu, t=t_0)$. We calculate the modified $z$-score of the quantity $\delta(\nu,t=t_0)=[S(\nu,t=t_0)-S_m(\nu,t=t_0)]$. Compared to regular z-score calculation, we use the modified z-score since it uses median absolute deviation ($MAD(x)=Median(|x_i-Median(x)|)$) and is less affected by extreme outliers. In our case it is calculated as follows: 
\begin{equation}
    z_m=\frac{0.675\times(\delta(\nu,t)-M[\delta(\nu,t)])}{\mathrm{MAD}}
\end{equation}

\begin{figure*}[!thb]
    \centering
    \includegraphics[width=0.9\linewidth]{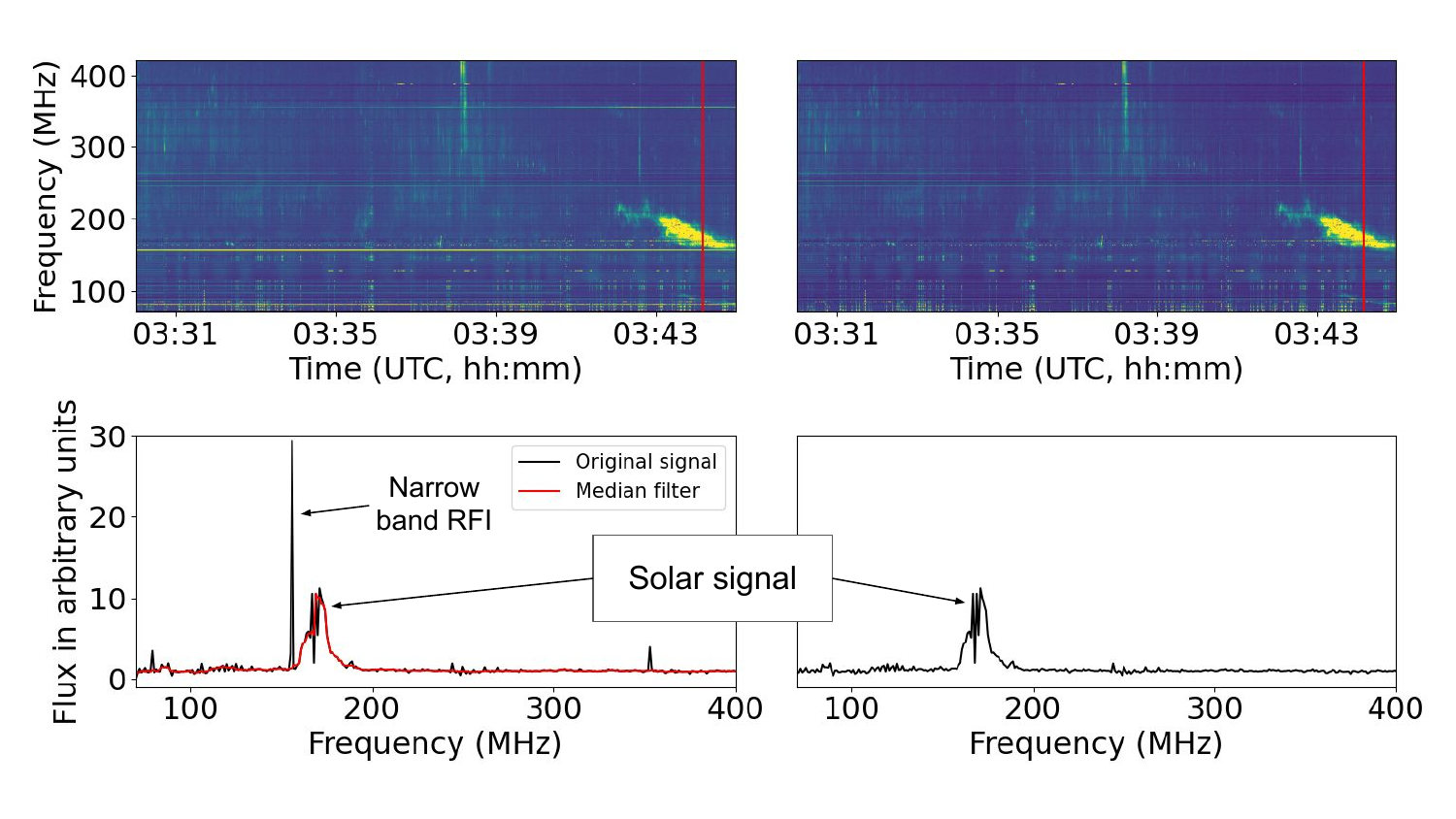}
    \caption{The top left and right panels show the dynamic spectrum before and after narrow band RFI removal, respectively. The bottom left and right panels show the same for spectra at UTC 03:44, as shown by the red vertical line on the dynamic spectrum in the top panel. The red trace in bottom left panel shows the median smoothed spectra ($S_m(\nu)$) of the same.}
    \label{fig:narrow}
\end{figure*}

Here, $M[\delta(\nu,t)]$ is calculated by applying a median filter to $\delta(\nu,t)$.  The factor of 0.675 comes from assuming that the distribution of $\delta(\nu,t)$ is a Gaussian. Any data point with a $z_m$-value more than 10 is replaced with the corresponding $S_m(\nu,t)$. The threshold of $z_m$ is chosen such that we avoid over-flagging of solar data.
The RFI-cleaned dynamic spectrum is then estimated as,
\begin{equation}
    S'(\nu,t)=
    \begin{cases}
    S(\nu,t),& \text{if } z_m<10\\
    S_m(\nu,t),              & \text{otherwise}
\end{cases}
\label{eq:rfi_cleaned_DS}
\end{equation}
Figure \ref{fig:narrow} shows the dynamic spectrum before and after narrowband RFI removal. As seen in the bottom panel, this method can filter out narrowband RFI even in presence of solar signal.

\subsection{Background Subtraction}\label{subsec:bkg_subtraction}
After bandpass normalization and RFI removal, the background has to be subtracted to enhance any transient feature in the dynamic spectra. The background is slowly varying and comes from instrumental gain variation, changes in system temperature, and the quiet Sun. 
To account for this variability, we need adaptive noise estimates rather than a constant background. The flux density distribution of a single channel for a sufficiently long time interval follows a Gaussian distribution. So, the median of this distribution represents the background level. On the time series for every spectral channel, we perform the following operation:
\begin{equation}
    S_{\mathrm{bksub}}(\nu=\nu_0,t)=S'(\nu=\nu_0,t)-M_T[S'(\nu=\nu_0,t)]
\end{equation}
Here, $M_T[S'(\nu=\nu_0,t)]$ represents the moving median of pixel values from timestamp $t-T$ to $t$ for frequency $\nu_0$. 
\begin{equation}
    M_T[S'(\nu_0,t)]=\mathrm{median}(S'(\nu_0,t-T),S'(\nu_0,t-T+1)...S'(\nu_0,t))
\end{equation}
To get good estimates of the background noise, we stitch together two consecutive 30-minute data chunks and calculate the median by setting the window width, $T$, to 1800 s. 
This duration is significantly larger than the lifetime of most solar radio bursts at any given frequency, and has been chosen with the motivation of reducing the likelihood of any solar radio bursts present in the window impacting the median value.

Finally, this background-subtracted dynamic spectrum, $S_\mathrm{bkgsub}(\nu,t)$, is used for further analysis. The top panels of figure \ref{fig:binary} show the dynamic spectrum before and after the median based background subtraction. As it can be clearly seen from the figure that the solar event has become much prominent.

\section{Event Detection}\label{sec:event_detection}
\subsection{Generating Binary Dynamic Spectrum}
The temporal resolution of the data received from the Yamagawa is 1 second and hence, to avoid false detection due to the presence of any low-level transient RFI, we only search for events which are present for more than at least 3 s. For each spectral channel, we estimate the background levels by calculating the standard deviation ($\sigma(\nu)$) from the previous day quiet time dynamic spectrum. A signal-to-noise ratio (SNR) dynamic spectrum of the real-time data is created after the following operation on each spectral channel --
\begin{equation}
    \text{SNR}=\frac{S_\mathrm{bkgsub}(\nu,t) }{\sigma(\nu)}
    \label{eq: snr_cal}
\end{equation}

We then make a binary dynamic spectrum by choosing pixels having values above a certain threshold. We find that a threshold of $13\sigma$ provides an optimized balance between false detection and missing true events, which is discussed later in section \ref{sec:performance}. 
\begin{figure*}[!thb]
    \centering
    \includegraphics[width=0.9\linewidth]{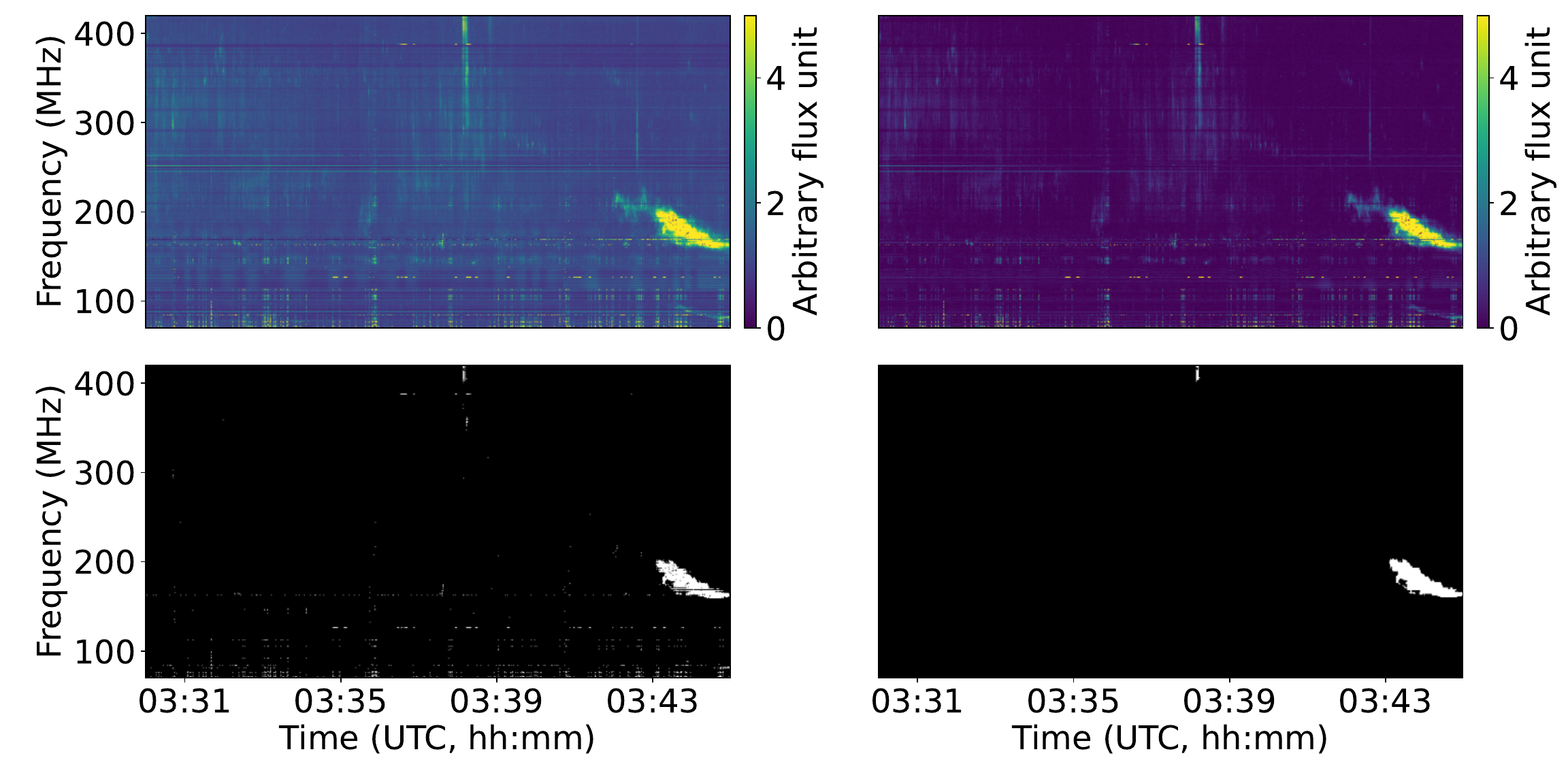}
    \caption{Top panels: the dynamic spectrum before and after the background subtraction. Bottom panels: (left) the binary dynamic spectrum  and (right) the detected closed contour region after morphological closing. The x-axes are shared across subplots in each column. The y-axes are shared across subplots in each row.}
    \label{fig:binary}
\end{figure*}
\subsection{Event Detection Algorithm}
To detect bursts, we look for isolated regions bound by closed contours in the binary dynamic spectrum. However, there can be missing pixels in binary dynamic spectrum due to RFI-flagging, which can make it harder to find closed contours following a simple approach. We perform a morphological closing operation \citep{serra1983image} on the binary dynamic spectrum to fill up any missing pixels inside it. This operation preserves the outer edge of any feature in binary image while filling up missing pixels inside it. The morphological closing is a 2 step process where the binary image is \textit{dilated} ($\oplus$) using the structuring kernel and then \textit{eroded} ($\ominus$) by the same kernel. The structuring kernel (K) is taken to be a 2$\times$2 matrix of ones. If the binary dynamic spectrum is represented by $S_{binary}$, we can write the operation as: 
\begin{equation}
    S_{closed}=(S_{binary}\oplus K)\ominus K
\end{equation}
We try to find closed contours on the image after morphological closing operation using \textit{OpenCV} which uses {\it border following method} \citep{suzuki1985topological}. Contours having minimum span of 15 MHz in frequency and 3 s in time are considered to be true events. This method not only finds closed contours but also filters out residual RFI which do not match the frequency and temporal span criteria. The bottom panels in figure \ref{fig:binary} show the detection of such a closed contour region in the binary dynamic spectrum. If we find any such region within the most recent two minutes, we consider this event to be observed by triggering the MWA.

\section{Triggering the MWA}\label{sec:mwa_triggering}
The final step is to communicate with the MWA once a triggering condition is met and to schedule appropriate solar observations. This is achieved via a web-based triggering mechanism that has been developed for the MWA \citep{Hancock}.

\begin{figure}
    \centering
    \includegraphics[width=\linewidth]{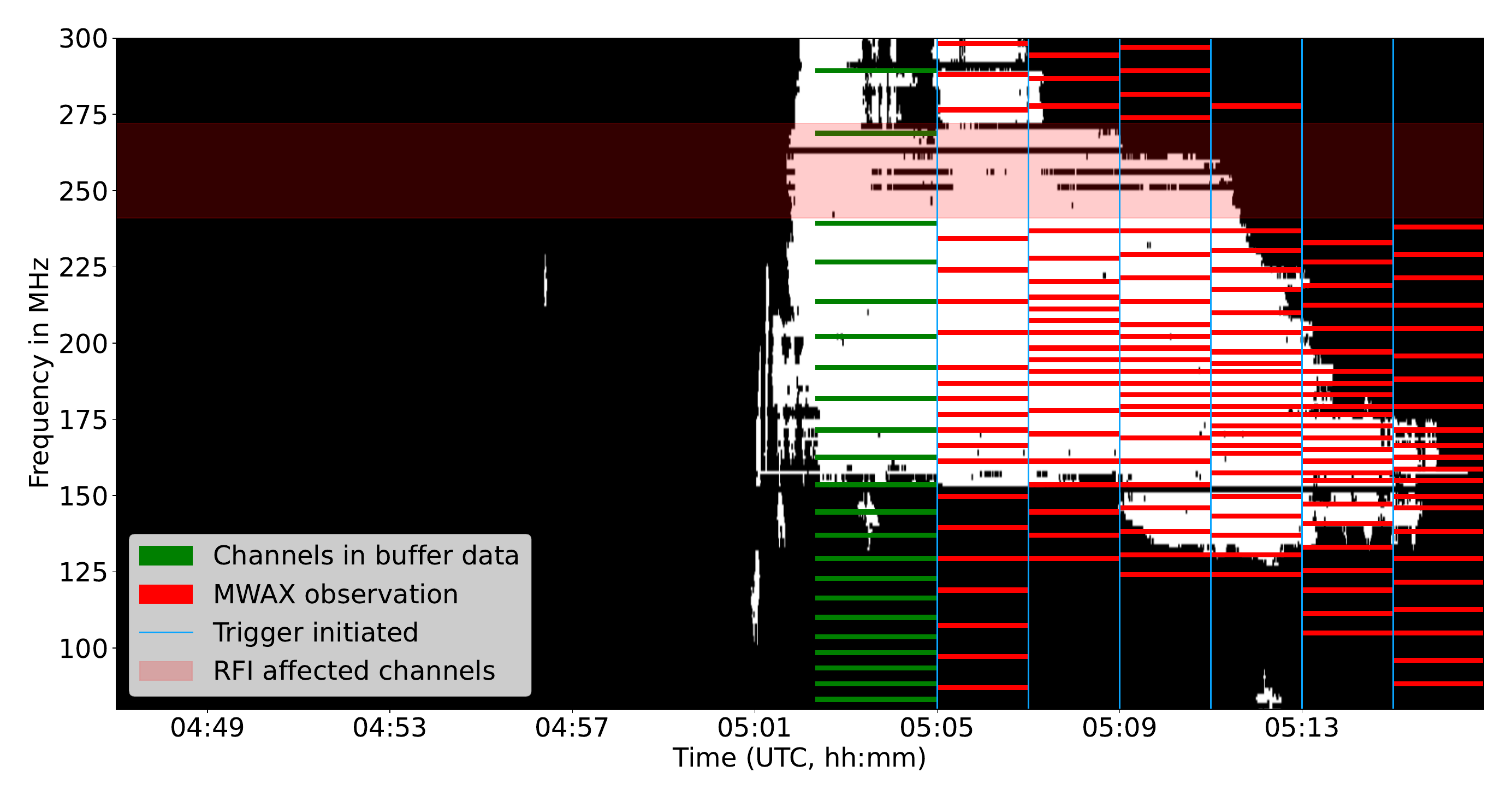}
    \caption{This figure shows how STORMY schedules subsequent MWAX observations after detection. The green rectangles are buffer channels. When STORMY triggers an observation (denoted by first blue vertical line), the 30.72 MHz band width is allocated according to the detected spectral span (red rectangles). In case of large bursts, as presented here, STORMY continues to adjust the frequency coverage as the burst drifts into lower frequency. The red transparent rectangle shows the RFI contaminated channels in MWA and are always excluded from observation.}
    \label{fig:triggered_ds}
\end{figure}

On arrival of a trigger, the buffer data is frozen and dumped to disk.
As mentioned in Sec. \ref{sec:instruments}, the buffer data is in `picket-fence' mode.  
Simultaneously, an interferometric observation for the next 2 minutes is initiated. STORMY's detection algorithm estimates the spectral span of the detected event. 
This information is used for defining the spectral coverage of the triggered observations. If $\nu_{min}$ and $\nu_{max}$ are the minimum and maximum detected frequency, the MWA observing bandwidth of 30.72 MHz is distributed linearly across $(\nu_{min}-30)$ MHz to $(\nu_{max}+10)$ MHz, subject to the MWA observing band of $80$--$300\ MHz$. It also avoids the RFI afflicted channels in frequency range of 241--271 MHz while determining the channels for observation.
The 30 MHz in lower limit and 10 MHz in upper limit are guard bands which allow for spectral drifts and the presence of weaker active emissions in the neighboring spectral regions which might not have been detectable at Yamagawa, but might be detectable by the more sensitive MWA.
The $30\ MHz$ guard band on the lower frequency side is chosen to accommodate the typical drift rate of type II bursts of $-0.1\ MHz\ s^{-1}$. Figure \ref{fig:triggered_ds} shows how STORMY chooses the spectral channels for interferometric observations. In case of long duration bursts, as shown in the figure, STORMY continues to detect and adjust the frequency coverage for subsequent interferometric observations.

\section{Quantifying the performance of STORMY}
\label{sec:performance}
Since, we will be triggering and observing based on the efficiency of the detection algorithm, it is important to get a quantitative sense for the accuracy and sensitivity of the trigger. To assess the performance of the pipeline, we use archival datasets with solar activity. We calculate the following for a range of SNR thresholds mentioned in Section \ref{sec:event_detection} : (1) Correct detection or true positives ($T_P$), (2) Incorrect detection or false positives ($F_P$) and (3) Incorrect non detection or false negatives ($F_N$).  These quantities are used to calculate the following metrics as described in \citet{fawcett2006}:
\begin{itemize}
    \item Precision: $T_P/(T_P+F_p)$, which measures the fraction of the total detections corresponding to true detections and hence quantifies how accurately the algorithm can detect true events.
    \item Recall: $T_P/(T_P+F_N)$, which measures the fraction of total true which events are correctly identified and hence quantifies how sensitively the algorithm can detect true events. 
    \item F1-score: 2$\times $Precision $\times $Recall/(Precision+Recall), represents the harmonic mean of precision and recall. An optimum F1 score is a balance between precision and recall.
\end{itemize}

\begin{figure}[!htbp]
    \centering
    \includegraphics[width=\linewidth]{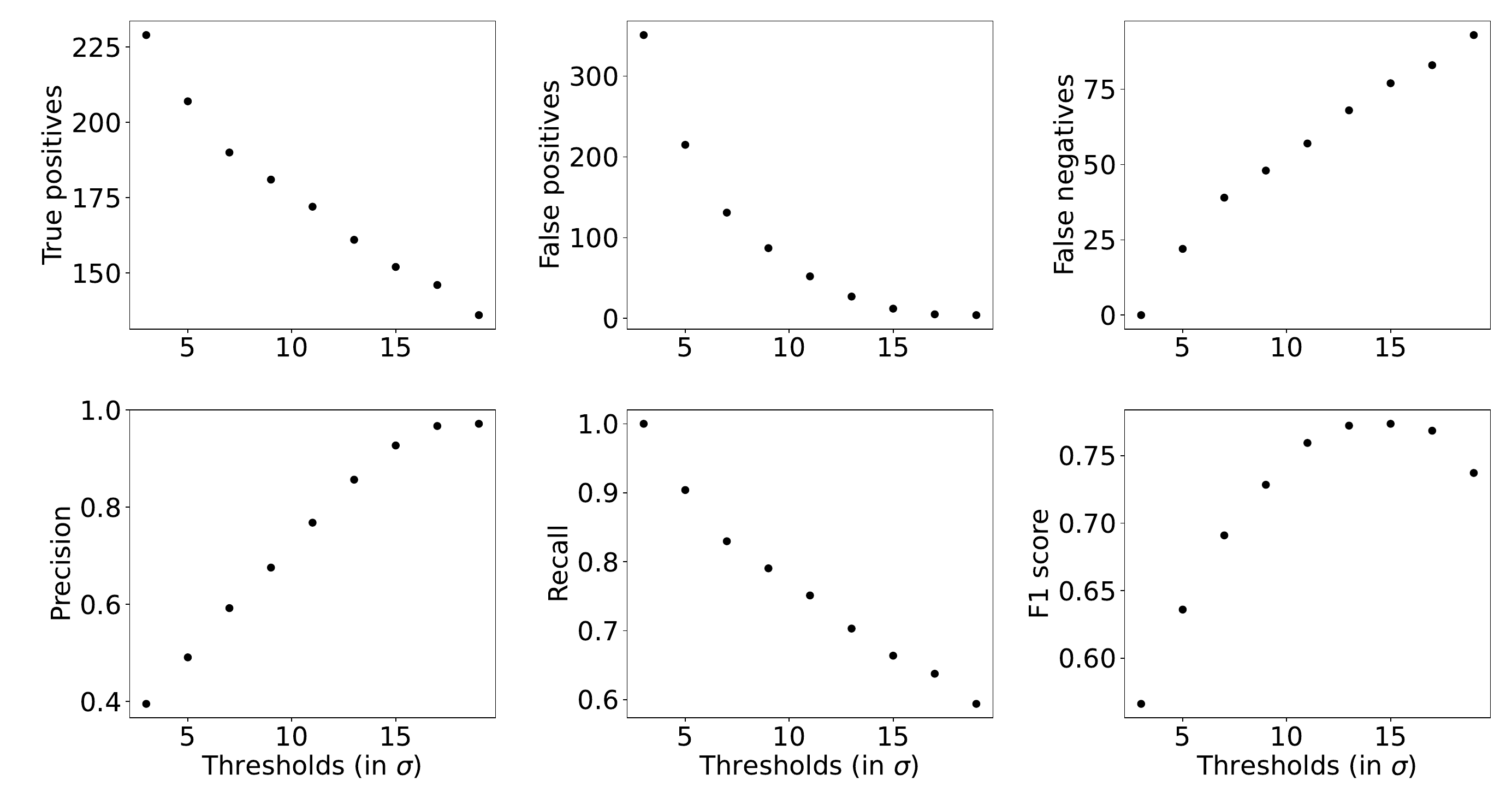}
    \caption{The figure shows various performance parameters tested for different threshold values. As can be seen from the plot that using thresholds around 13$\sigma$ gives the best F1-score.}
    \label{fig:performance}
\end{figure}

Since, we trigger on 2 minutes chunks, we tested our algorithm for different thresholds on an ensemble of 2-minute templates. We have used 8400 such 2-minute chunks, where 226 true events were present. Figure \ref{fig:performance} shows the results from the performance test. From the F1-score plot we judge that using threshold in the range of 10-15$\sigma$ gives the best precision, recall pair.

\section{Some Sample Events Captured by STORMY}\label{sec:final_results}The triggering pipeline, STORMY, has been made operational at the MWA system since around August 2024. We have successfully triggered and observed $\sim110$ bursts in the period from 31st July 2024 to 31st January 2025, showcasing the capability of the trigger in capturing event-rich data. Imaging of the interferometric and buffer data shows promising results. Here, we present two sample events observed using STORMY triggers --
\begin{itemize}
    \item \textbf{Event 1:} On 1st August 2024, around 1:47 UTC a GOES M-class X-ray flare took place at the active region (AR) NOAA 13773. A bright solar radio burst was detected and observations were triggered using STORMY. The top left panel of Figure \ref{fig:1st_aug} shows the Yamagawa dynamic spectra . We made a quick-look image using the triggered observation. The top right panel of figure \ref{fig:1st_aug} shows the radio image at 126 MHz. The bottom panels show the evolution of the event as seen by SDO/AIA in 171\AA.

\begin{figure*}[!thb]
    \centering
    \includegraphics[width=\linewidth]{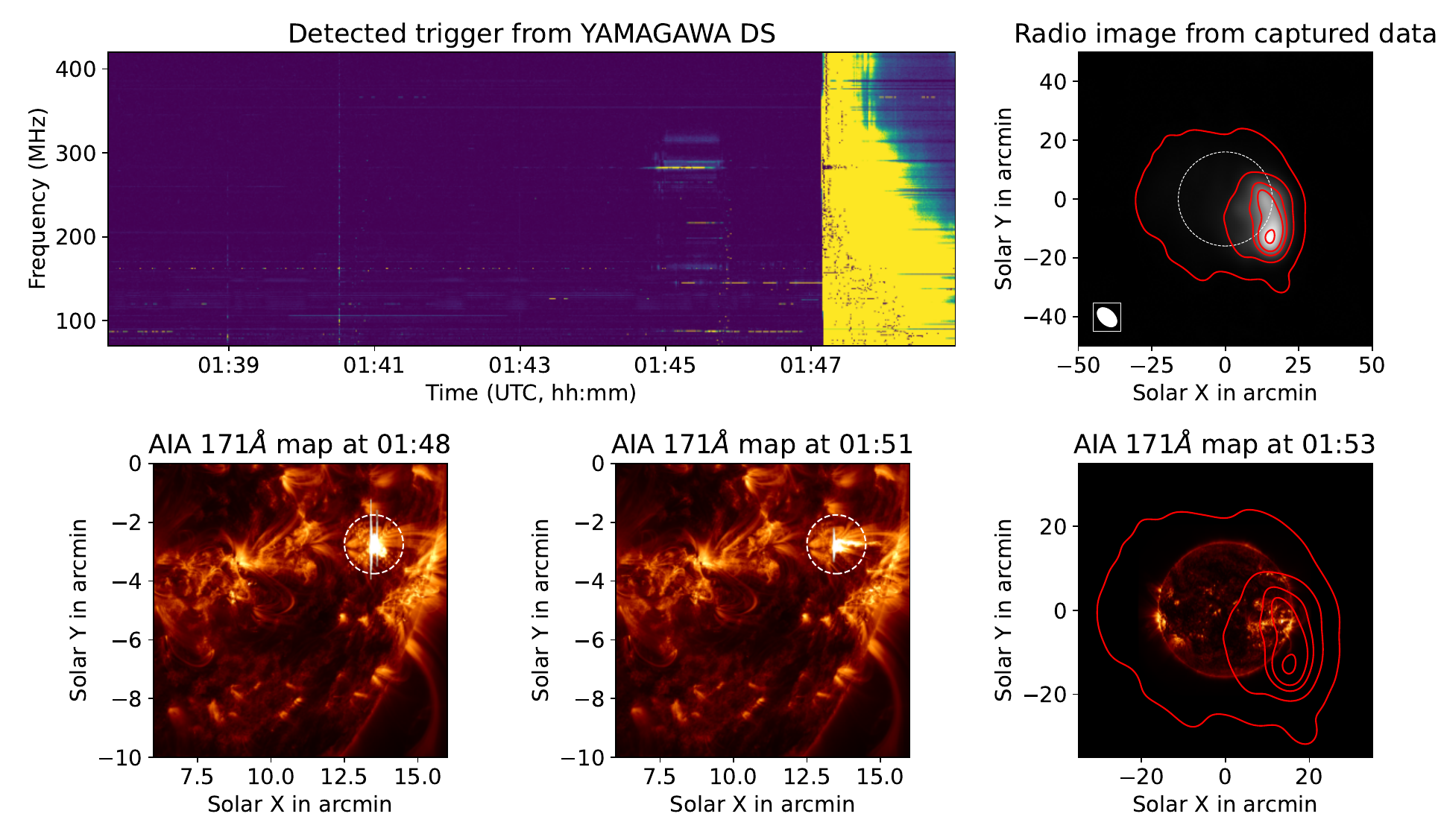}
    \caption{Top panels: (Left) Pre-processed Yamagawa dynamic spectrum. (Right) Radio image of the data captured at 126 MHz on 1 August 2024.The dotted white circle shows the optical disc of the Sun. The red contours are at the 1,10,30,50,90\% of the peak. The ellipse inside the box at the bottom left corner shows the point-spread-function. Bottom panels: (Left and Middle) close-up of AIA images during the event. The dotted circles show the active region from which the eruption is taking place. (Right) Radio contours overlaid on the AIA map. }
    \label{fig:1st_aug}
\end{figure*}
\begin{figure*}[!thb]
    \centering
    \includegraphics[width=\linewidth]{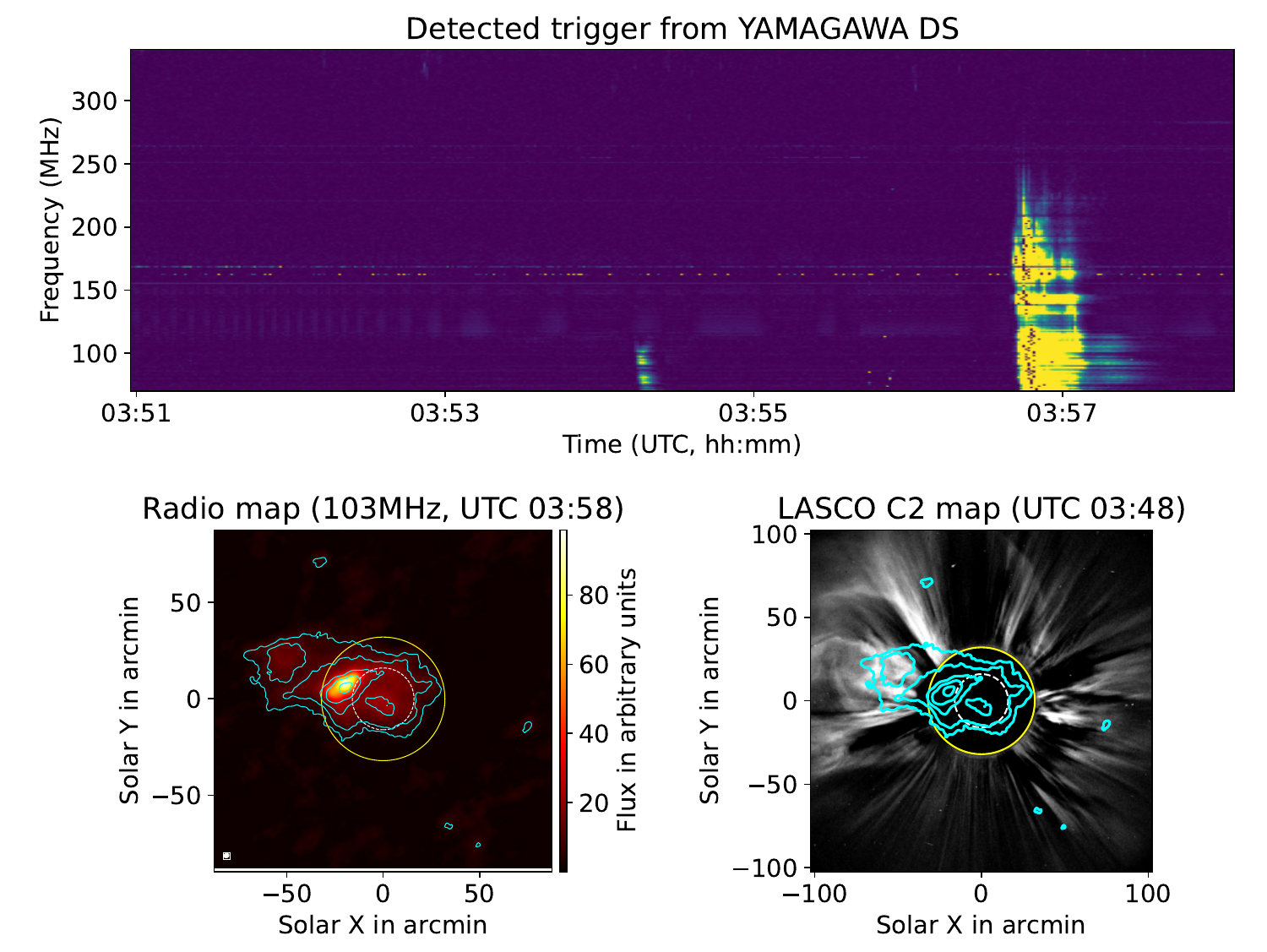}
    \caption{(Top panel) Pre-processed Yamagawa dynamic spectrum. (Bottom left) Radio image from buffer data of the MWA at 103 MHz taken through triggered observation on 4 November 2024. The dotted white circle shows the optical disk of the Sun and the solid yellow circle has radius of 2$R_\odot$. The cyan contours are at levels 5,10,20,40,80\% of the peak. The white ellipse inside the rectangle box is the psf. (Bottom right) Radio contours overlaid on LASCO C2 base difference image.}
    \label{fig:4th_nov}
\end{figure*}
    \item \textbf{Event 2:} On 4 November 2024, STORMY triggered on a group type-III solar event. The images, obtained from the buffer data, are very promising. In figure \ref{fig:4th_nov} we show a radio image made using the buffer data at 103 MHz. The radio contours were overlaid on LASCO C2 base difference image. 
\end{itemize}

\section{Discussion and Conclusion}\label{sec:application}
Though developed with the objective of increasing the observing efficiency of solar activity by the MWA, the framework of STORMY has wider applicability. Next, we discuss some of the few specific instances as examples where solar triggering can be implemented on modern radio interferometers:

\begin{itemize}
    \item \textbf{LOFAR: } The \textbf{LO}w \textbf{F}requency \textbf{AR}rray is a versatile instrument, primarily due to its fully digital signal chain \citep{lofar_paper}. It can observe multiple sources simultaneously via its multi-beam and subarray capabilities. The upcoming LOFAR 2.0 \citep{lofar2023whitepaper} includes the LOFAR for Space Weather (LOFAR4SW) extension, which is planned to use few stations to track the Sun while it is above the instrument's elevation limit commensal with other astronomical observations, simultaneously for the Low-Band Antennas (LBA; 10--90 MHz) and High-Band Antennas (HBA; 110--240 MHz). 
    Using a part of the same instrument for solar monitoring offers several advantages over using a remote spectrograph - (a) the trigger comes from a monitoring system of better sensitivity and experiences RFI environment similar to the observing system, (b) co-location of the monitoring station removes the dependency on an independent distant system and eliminates the delays arising from the time taken to fetch remote data, substantially reducing the memory requirement of the ring buffer. 
    Additionally, in LOFAR 2.0, having the ability to monitor the Sun simultaneously in HBA and LBA, can potentially enable one to raise an observing trigger for the LBA just a little before the emission drifts into the observing band of LBA.
    As an illustration, we demonstrate an application of STORMY to LOFAR data in figure \ref{fig:lofar}. 
    These data come from the LBA beam-formed mode, which was recorded with a spectral resolution of 12.207 kHz and temporal resolution of 10.486 ms, and have been averaged to 195.3 kHz and 1 s for faster processing.
    STORMY's RFI excision and burst detection algorithms are seen to work very well on the LOFAR dynamic spectra. It took STORMY $\sim$30 $s$ to process the whole 40-minute datachunk. Hence, STORMY can be used on real-time beam-formed data to detect events and trigger interferometric observation when appropriate.
    
   \begin{figure*}[!thb]
    \includegraphics[width=\linewidth]{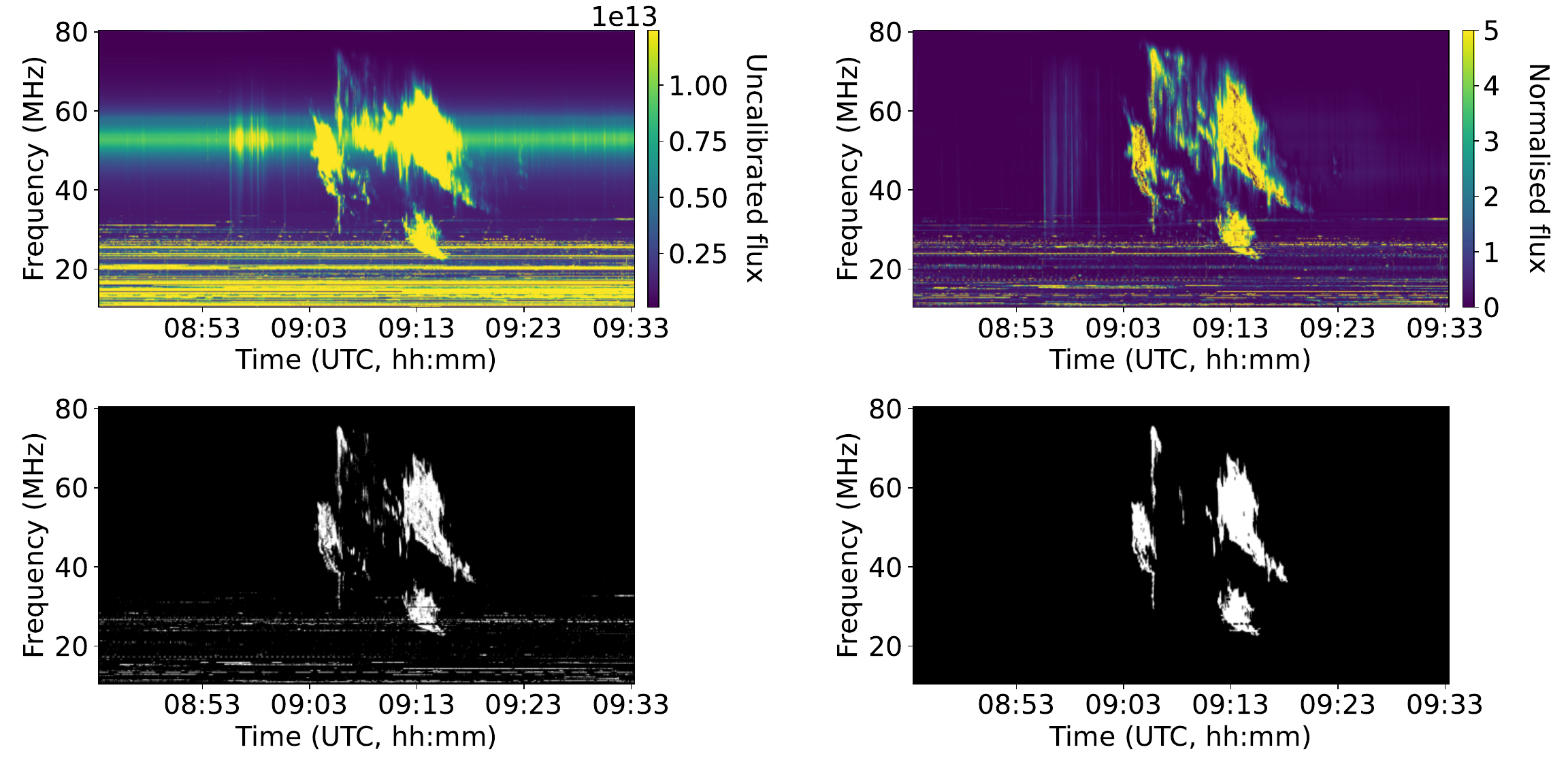}
    \caption{This figure shows the LOFAR dynamic spectrum processed through the steps mentioned before. The top left panel shows the raw beam formed data from LOFAR. The top right panel shows the pre-processed data. The bottom left panel shows the binary dynamic spectrum after using 5$\sigma$ threshold. The bottom right panel shows the detected closed contour regions from the binary dynamic spectrum.}
    \label{fig:lofar}
    \end{figure*}
    \item \textbf{SKAO: } The triggering aspects are particularly relevant for the SKA-Low, which operates in the 50--350 MHz band where the bulk of solar radio activity is observed. Like the MWA, it uses aperture array elements and hence electronic pointing, essentially eliminating the slew time needed to point to the target source, post the trigger. 
    The SKA-Low is expected to reach its AA* milestone before the end of this decade \citep{SKAO-YearInLife}. 
    It will have 307 stations, each comprising 256 dual-polarization dipoles log-periodic antennas placed over a footprint of $38\ m$ diameter.  
    The SKA-Low supports flexible options for forming not only subarrays, but also substations, with the latter providing access to larger FoV as well as shorter baselines (within a station), though subject to bandwidth restrictions imposed by the correlator and the network capacity. 
    AA* will support up to 1440 substations. 
    The smallest substations have a $6\ m$ foot print and typically include 6 dual-pol antennas \citep{SKAO-Low-SubStation-Templates}.
    A subarray comprising one such substation from $\sim$60 SKA-Low stations within a diameter of $\sim10\ km$ will be an excellent match to the needs of triggered solar observing and is a referred to as {\em solar subarray} in the following text.
    It uses $< 0.5$\% of the SKA-Low collecting area, and hence has minimal sensitivity impact on the observations it is commensal with.
    Along the lines of the picket fence mode routinely employed at the MWA for solar observations, the solar subarray can use narrow spectral channels distributed across the 50--350 MHz SKA-Low observing band.
    For instance, using a hundred spectral channels, each 5.4 kHz in bandwidth, amounts to a mere 0.54 MHz of bandwidth.
    The solar subarray can be used to fill the voltage buffer while simultaneously being analyzed for the presence of solar activity to raise a trigger. SKA-Low has a substantial transient buffer, spanning 900~$s$ \citep{SKAO-YearInLife}. To enable the capability of retrieving the event data lost due to time taken for trigger detection and generation, the solar subarray data can be directed to the buffer. 
    On the arrival of a trigger the voltage buffer can be saved to disk and interferometric observations can be commenced with a denser subarray and using a larger observing bandwidth. \\
    Other options, which use small subarrays to monitor the Sun with tied array beams, have also been suggested \citep{seethapuram_sridhar_2025_16951088}. 
    They can provide data which can serve as inputs to STORMY's near-real time solar activity detection module and trigger SKA observations on detection of suitable solar activity. 
    Some of these options are tailored for SKA-Mid and suggest the use a set of 3 subarrays, of 4 antenna each, to simultaneously monitor solar activity in 3 SKA-mid bands using a STORMY-like approach and raise observing triggers as appropriate.

    \item \textbf{Solar dedicated instruments: } 
    An interesting use case of STORMY for some of the new Solar-dedicated instruments is that the triggers can be used to identify the temporal and spectral span of Solar events in near-real time. This information can, for instance, be used for switching to a mode which saves visibilities at higher time and/or frequency resolutions or perhaps optimizing other observing parameters as well. 
    Such a usage is actively under consideration at the OVRO-LWA.

\end{itemize}

In conclusion, our triggering framework, STORMY, provides a robust and efficient means to capture interferometric data of transient solar activity. We also show how such framework can be implemented in other instruments to substantially increase the efficiency of gathering interesting data needed for investigating the Sun at radio frequencies. 

\section{Acknowledgement}
This scientific work uses data obtained from Inyarrimanha Ilgari Bundara, the CSIRO Murchison Radio-astronomy Observatory. We acknowledge the Wajarri Yamaji People as the Traditional Owners and Native Title Holders of the observatory site. Support for the operation of the MWA is provided by the Australian Government (NCRIS), under a contract to Curtin University administered by Astronomy Australia Limited. This work was supported by resources provided by the Pawsey Supercomputing Research Centre’s \href{https://doi.org/10.48569/18sb-8s43}{ Setonix Supercomputer} and \href{https://doi.org/10.48569/gskb-tp15}{Garrawarla GPU Cluster}. We thank the National Institute of Information and Communications Technology (NICT) for providing the Yamagawa data. We sincerely thank Kazumasa Iwai (Nagoya University) for helping us to initiate conversation with NICT. We acknowledge Pietro Zucca for providing the LOFAR beam-formed data used in this work. D. P., S. D. and D. O. acknowledge the support of the Department of Atomic Energy, Government of India, under project no. 12-R\&D-TFR-5.02-0700. D. K. acknowledges the support by the NASA Living with a Star Jack Eddy Postdoctoral Fellowship Program, administered by UCAR’s Cooperative Programs for the Advancement of Earth System Science (CPAESS) under award 80NSSC22M0097. We thank Barnali Das (CSIRO, Australia) for providing the name of the pipeline and the anomymous referees for constructive comments which improved the clarity and presentation of this work.

\bibliography{sola_bibliography_example}  

\begin{thebibliography}{}
\expandafter\ifx\csname natexlab\endcsname\relax\def\natexlab#1{#1}\fi

\bibitem[{{Altyntsev} {et~al.}(2020){Altyntsev}, {Lesovoi}, {Globa}, {Gubin}, {Kochanov}, {Grechnev}, {Ivanov}, {Kobets}, {Meshalkina}, {Muratov}, {Prosovetsky}, {Myshyakov}, {Uralov}, \& {Fedotova}}]{srh_2020}
{Altyntsev}, A., {Lesovoi}, S., {Globa}, M., {et~al.} 2020, Solar-Terrestrial Physics, 6, 30

\bibitem[{{Anderson} {et~al.}(2024){Anderson}, {Schroeder}, {van der Horst}, {Rhodes}, {Rowlinson}, {Bahramian}, {Chastain}, {Gompertz}, {Hancock}, {Laskar}, {Leung}, \& {Wijers}}]{grb_triggered_2024}
{Anderson}, G.~E., {Schroeder}, G., {van der Horst}, A.~J., {et~al.} 2024, \apjl, 975, L13

\bibitem[{Bastian {et~al.}(1998)Bastian, Benz, \& Gary}]{bastian1998radio}
Bastian, T., Benz, A., \& Gary, D. 1998, Annual Review of Astronomy and Astrophysics, 36, 131

\bibitem[{Benz {et~al.}(2009)Benz, Monstein, Meyer, Manoharan, Ramesh, Altyntsev, Lara, Paez, \& Cho}]{benz2009world}
Benz, A., Monstein, C., Meyer, H., {et~al.} 2009, Earth, Moon, and Planets, 104, 277

\bibitem[{Benz {et~al.}(2005)Benz, Monstein, \& Meyer}]{benz2005callisto}
Benz, A.~O., Monstein, C., \& Meyer, H. 2005, Solar Physics, 226, 143

\bibitem[{{Bhunia} {et~al.}(2023){Bhunia}, {Carley}, {Oberoi}, \& {Gallagher}}]{shilpi_type2}
{Bhunia}, S., {Carley}, E.~P., {Oberoi}, D., \& {Gallagher}, P.~T. 2023, \aap, 670, A169

\bibitem[{Boischot {et~al.}(1980)Boischot, Rosolen, Aubier, Daigne, Genova, Leblanc, Lecacheux, de~La~Noe, {et~al.}}]{boischot1980new}
Boischot, A., Rosolen, C., Aubier, M., {et~al.} 1980, Icarus, 43, 399

\bibitem[{Bonnin {et~al.}(2011)Bonnin, Aboudarham, Fuller, Renie, Perez-Suarez, Gallagher, Higgins, Krista, Csillaghy, \& Bentley}]{bonnin2011automated}
Bonnin, X., Aboudarham, J., Fuller, N., {et~al.} 2011, in SF2A-2011: Proc. Annu. Meeting French Soc. Astron. Astrophys, Vol. 373, 377

\bibitem[{Chhabra {et~al.}(2021)Chhabra, Gary, Hallinan, Anderson, Chen, Greenhill, \& Price}]{chhabra2021imaging}
Chhabra, S., Gary, D.~E., Hallinan, G., {et~al.} 2021, The Astrophysical Journal, 906, 132

\bibitem[{Cla{\ss}en \& Aurass(2002)}]{classen2002association}
Cla{\ss}en, H.-T., \& Aurass, H. 2002, Astronomy \& Astrophysics, 384, 1098

\bibitem[{Deng {et~al.}(2024)Deng, Yuan, Zhou, Wu, \& Tan}]{deng2024real}
Deng, J., Yuan, G., Zhou, H., Wu, H., \& Tan, C. 2024, Astrophysics and Space Science, 369, 99

\bibitem[{Dey {et~al.}(2025)Dey, Kansabanik, Oberoi, \& Mondal}]{dey2025first}
Dey, S., Kansabanik, D., Oberoi, D., \& Mondal, S. 2025, The Astrophysical Journal Letters, 988, L73

\bibitem[{{Dey} {et~al.}(2025){Dey}, {Oberoi}, {Zucca}, {Mancini}, {Patra}, \& {Kansabanik}}]{dey2025automated}
{Dey}, S., {Oberoi}, D., {Zucca}, P., {et~al.} 2025, Astronomy \& Astrophysics, 704, A75

\bibitem[{Fawcett(2006)}]{fawcett2006}
Fawcett, T. 2006, Pattern recognition letters, 27, 861

\bibitem[{Gary(2023)}]{gary2023new}
Gary, D.~E. 2023, Annual Review of Astronomy and Astrophysics, 61, 427

\bibitem[{Gary {et~al.}(2018)Gary, Chen, Dennis, Fleishman, Hurford, Krucker, McTiernan, Nita, Shih, White, {et~al.}}]{gary2018microwave}
Gary, D.~E., Chen, B., Dennis, B.~R., {et~al.} 2018, The Astrophysical Journal, 863, 83

\bibitem[{Gergely \& Mahoney(1985)}]{Gregley1985}
Gergely, T.~E., \& Mahoney, M.~J. 1985, 411

\bibitem[{Ginzburg \& Zhelezniakov(1959)}]{ginzburg1959mechanisms}
Ginzburg, V., \& Zhelezniakov, V. 1959, in Symposium-International Astronomical Union, Vol.~9, Cambridge University Press, 574--582

\bibitem[{Gopalswamy {et~al.}(2005)Gopalswamy, Aguilar-Rodriguez, Yashiro, Nunes, Kaiser, \& Howard}]{gopalswamy2005type}
Gopalswamy, N., Aguilar-Rodriguez, E., Yashiro, S., {et~al.} 2005, Journal of Geophysical Research: Space Physics, 110

\bibitem[{Hancock {et~al.}(2019)Hancock, Anderson, Williams, Sokolowski, Tremblay, Rowlinson, Crosse, Meyers, Lynch, Zic, \& et~al.}]{Hancock}
Hancock, P.~J., Anderson, G.~E., Williams, A., {et~al.} 2019, Publications of the Astronomical Society of Australia, 36, e046

\bibitem[{Hough(1962)}]{hough1962}
Hough, P. V.~C. 1962, Method and means for recognizing complex patterns

\bibitem[{Iwai {et~al.}(2017)Iwai, Kubo, Ishibashi, Naoi, Harada, Ema, Hayashi, \& Chikahiro}]{Iwai_2017}
Iwai, K., Kubo, Y., Ishibashi, H., {et~al.} 2017, Earth, Planets and Space, 69, doi:10.1186/s40623-017-0681-8

\bibitem[{{Jonas} \& {MeerKAT Team}(2016)}]{meerkat_2016}
{Jonas}, J., \& {MeerKAT Team}. 2016, in MeerKAT Science: On the Pathway to the SKA, 1

\bibitem[{Kansabanik(2022)}]{Kansabanik2022c}
Kansabanik, D. 2022, Solar Physics, 297, 122

\bibitem[{Kansabanik {et~al.}(2023)Kansabanik, Bera, Oberoi, \& Mondal}]{Kansabanik_2023}
Kansabanik, D., Bera, A., Oberoi, D., \& Mondal, S. 2023, The Astrophysical Journal Supplement Series, 264, 47

\bibitem[{Kansabanik {et~al.}(2024)Kansabanik, Mondal, \& Oberoi}]{kansabanik2024spectropolarimetric}
Kansabanik, D., Mondal, S., \& Oberoi, D. 2024, The Astrophysical Journal, 968, 55

\bibitem[{Kansabanik {et~al.}(2022)Kansabanik, Oberoi, \& Mondal}]{Kansabanik_2022}
Kansabanik, D., Oberoi, D., \& Mondal, S. 2022, The Astrophysical Journal, 932, 110

\bibitem[{{Kansabanik} {et~al.}(2025){Kansabanik}, {Gouws}, {Patra}, {Vourlidas}, {Kotz{\'e}}, {Oberoi}, {Shaik}, {Buchner}, \& {Camilo}}]{kansabanik2025solar}
{Kansabanik}, D., {Gouws}, M., {Patra}, D., {et~al.} 2025, Frontiers in Astronomy and Space Sciences, 12, 1666743

\bibitem[{{Kerdraon} \& {Delouis}(1997)}]{kerdraon}
{Kerdraon}, A., \& {Delouis}, J.-M. 1997, in Coronal Physics from Radio and Space Observations, ed. G.~{Trottet}, Vol. 483, 192

\bibitem[{{Lobzin} {et~al.}(2010){Lobzin}, {Cairns}, {Robinson}, {Steward}, \& {Patterson}}]{Lobzinn}
{Lobzin}, V.~V., {Cairns}, I.~H., {Robinson}, P.~A., {Steward}, G., \& {Patterson}, G. 2010, \apjl, 710, L58

\bibitem[{{LOFAR2.0 White Paper}(2023)}]{lofar2023whitepaper}
{LOFAR2.0 White Paper}. 2023, LOFAR2.0 White Paper -- v2023.1: A premier low-frequency radio telescope for the 2020s, Tech. rep., International LOFAR Telescope (ILT), first printing, April 2023. Licensed under Creative Commons Attribution-NonCommercial 3.0 Unported License. To be published on arXiv.

\bibitem[{{Mann} {et~al.}(2003){Mann}, {Klassen}, {Aurass}, \& {Classen}}]{mann_formation}
{Mann}, G., {Klassen}, A., {Aurass}, H., \& {Classen}, H.~T. 2003, \aap, 400, 329

\bibitem[{{Marcote} {et~al.}(2020){Marcote}, {Nimmo}, {Hessels}, {Tendulkar}, {Bassa}, {Paragi}, {Keimpema}, {Bhardwaj}, {Karuppusamy}, {Kaspi}, {Law}, {Michilli}, {Aggarwal}, {Andersen}, {Archibald}, {Bandura}, {Bower}, {Boyle}, {Brar}, {Burke-Spolaor}, {Butler}, {Cassanelli}, {Chawla}, {Demorest}, {Dobbs}, {Fonseca}, {Giri}, {Good}, {Gourdji}, {Josephy}, {Kirichenko}, {Kirsten}, {Landecker}, {Lang}, {Lazio}, {Li}, {Lin}, {Linford}, {Masui}, {Mena-Parra}, {Naidu}, {Ng}, {Patel}, {Pen}, {Pleunis}, {Rafiei-Ravandi}, {Rahman}, {Renard}, {Scholz}, {Siegel}, {Smith}, {Stairs}, {Vanderlinde}, \& {Zwaniga}}]{frb_trigger_2020}
{Marcote}, B., {Nimmo}, K., {Hessels}, J.~W.~T., {et~al.} 2020, \nat, 577, 190

\bibitem[{{Maxwell} \& {Swarup}(1958)}]{maxwell_swarup}
{Maxwell}, A., \& {Swarup}, G. 1958, \nat, 181, 36

\bibitem[{McLean \& Labrum(1985)}]{mclean1985solar}
McLean, D.~J., \& Labrum, N.~R. 1985, Solar Radiophysics: Studies of Emission from the Sun at Metre Wavelengths

\bibitem[{Melrose(1980)}]{melrose1980emission}
Melrose, D. 1980, Space Science Reviews, 26, 3

\bibitem[{Melrose(2017)}]{melrose2017coherent}
---. 2017, Reviews of Modern Plasma Physics, 1, 1

\bibitem[{{Morosan} {et~al.}(2014){Morosan}, {Gallagher}, {Zucca}, {Fallows}, {Carley}, {Mann}, {Bisi}, {Kerdraon}, {Konovalenko}, {MacKinnon}, {Rucker}, {Thid{\'e}}, {Magdaleni{\'c}}, {Vocks}, {Reid}, {Anderson}, {Asgekar}, {Avruch}, {Bentum}, {Bernardi}, {Best}, {Bonafede}, {Bregman}, {Breitling}, {Broderick}, {Br{\"u}ggen}, {Butcher}, {Ciardi}, {Conway}, {de Gasperin}, {de Geus}, {Deller}, {Duscha}, {Eisl{\"o}ffel}, {Engels}, {Falcke}, {Ferrari}, {Frieswijk}, {Garrett}, {Grie{\ss}meier}, {Gunst}, {Hassall}, {Hessels}, {Hoeft}, {H{\"o}randel}, {Horneffer}, {Iacobelli}, {Juette}, {Karastergiou}, {Kondratiev}, {Kramer}, {Kuniyoshi}, {Kuper}, {Maat}, {Markoff}, {McKean}, {Mulcahy}, {Munk}, {Nelles}, {Norden}, {Orru}, {Paas}, {Pandey-Pommier}, {Pandey}, {Pietka}, {Pizzo}, {Polatidis}, {Reich}, {R{\"o}ttgering}, {Scaife}, {Schwarz}, {Serylak}, {Smirnov}, {Stappers}, {Stewart}, {Tagger}, {Tang}, {Tasse}, {Thoudam}, {Toribio}, {Vermeulen}, {van Weeren}, {Wucknitz}, {Yatawatta}, \& {Zarka}}]{morosan_2014}
{Morosan}, D.~E., {Gallagher}, P.~T., {Zucca}, P., {et~al.} 2014, \aap, 568, A67

\bibitem[{{Morrison} {et~al.}(2023){Morrison}, {Crosse}, {Sleap}, {Wayth}, {Williams}, {Johnston-Hollitt}, {Jones}, {Tingay}, {Walker}, \& {Williams}}]{morrison2023mwax}
{Morrison}, I.~S., {Crosse}, B., {Sleap}, G., {et~al.} 2023, \pasa, 40, e019

\bibitem[{{Nakajima} {et~al.}(1994){Nakajima}, {Nishio}, {Enome}, {Shibasaki}, {Takano}, {Hanaoka}, {Torii}, {Sekiguchi}, {Bushimata}, {Kawashima}, {Shinohara}, {Irimajiri}, {Koshiishi}, {Kosugi}, {Shiomi}, {Sawa}, \& {Kai}}]{nakajima}
{Nakajima}, H., {Nishio}, M., {Enome}, S., {et~al.} 1994, IEEE Proceedings, 82, 705

\bibitem[{Nindos(2020)}]{nindos2020incoherent}
Nindos, A. 2020, Frontiers in Astronomy and Space Sciences, 7, 57

\bibitem[{Oberoi {et~al.}(2023)Oberoi, Bisoi, Raja, Kansabanik, Mohan, Mondal, \& Sharma}]{oberoi2023preparing}
Oberoi, D., Bisoi, S.~K., Raja, K.~S., {et~al.} 2023, Journal of Astrophysics and Astronomy, 44, 40

\bibitem[{Oru{\'e} {et~al.}(2023)Oru{\'e}, Stalder, Salgueiro, \& Molina}]{orue2023automatic}
Oru{\'e}, I.~G., Stalder, D., Salgueiro, L., \& Molina, J. 2023, in 2023 IEEE CHILEAN Conference on Electrical, Electronics Engineering, Information and Communication Technologies (CHILECON), IEEE, 1--6

\bibitem[{{Ramesh} {et~al.}(1998){Ramesh}, {Subramanian}, {SundaraRajan}, \& {Sastry}}]{Ramesh1998}
{Ramesh}, R., {Subramanian}, K.~R., {SundaraRajan}, M.~S., \& {Sastry}, C.~V. 1998, \solphys, 181, 439

\bibitem[{{Reid} \& {Kontar}(2017)}]{reid_kontar}
{Reid}, H. A.~S., \& {Kontar}, E.~P. 2017, \aap, 606, A141

\bibitem[{Saint-Hilaire {et~al.}(2012)Saint-Hilaire, Vilmer, \& Kerdraon}]{saint2012decade}
Saint-Hilaire, P., Vilmer, N., \& Kerdraon, A. 2012, The Astrophysical Journal, 762, 60

\bibitem[{{Salmane} {et~al.}(2018){Salmane}, {Weber}, {Abed-Meraim}, {Klein}, \& {Bonnin}}]{salmane}
{Salmane}, H., {Weber}, R., {Abed-Meraim}, K., {Klein}, K.-L., \& {Bonnin}, X. 2018, Journal of Space Weather and Space Climate, 8, A43

\bibitem[{Scully {et~al.}(2021)Scully, Flynn, Carley, Gallagher, \& Daly}]{scully2021type}
Scully, J., Flynn, R., Carley, E., Gallagher, P., \& Daly, M. 2021, in 2021 32nd Irish Signals and Systems Conference (ISSC), IEEE, 1--6

\bibitem[{Serra(1983)}]{serra1983image}
Serra, J. 1983, Image Analysis and Mathematical Morphology

\bibitem[{Shibasaki(2013)}]{shibasaki2013long}
Shibasaki, K. 2013, Publications of the Astronomical Society of Japan, 65, S17

\bibitem[{{Singh} {et~al.}(2019){Singh}, {Sasikumar Raja}, {Subramanian}, {Ramesh}, \& {Monstein}}]{Dayal}
{Singh}, D., {Sasikumar Raja}, K., {Subramanian}, P., {Ramesh}, R., \& {Monstein}, C. 2019, \solphys, 294, 112

\bibitem[{Sridhar {et~al.}(2025{\natexlab{a}})Sridhar, Breen, Chrysostomou, \& Ball}]{SKAO-YearInLife}
Sridhar, S., Breen, S., Chrysostomou, A., \& Ball, L. 2025{\natexlab{a}}, A year in the life of the SKA telescopes: overview and main outcomes, doi:10.5281/zenodo.16950982

\bibitem[{Sridhar {et~al.}(2025{\natexlab{b}})Sridhar, Williams, Price, Breen, \& Ball}]{seethapuram_sridhar_2025_16951088}
Sridhar, S., Williams, W., Price, D., Breen, s., \& Ball, L. 2025{\natexlab{b}}, SKA Low and Mid subarray templates, doi:10.5281/zenodo.16951088

\bibitem[{Suzuki {et~al.}(1985)}]{suzuki1985topological}
Suzuki, S., {et~al.} 1985, Computer vision, graphics, and image processing, 30, 32

\bibitem[{Takano {et~al.}(2007)Takano, Nakajima, Enome, Shibasaki, Nishio, Hanaoka, Shiomi, Sekiguchi, Kawashima, Bushimata, {et~al.}}]{takano2007upgrade}
Takano, T., Nakajima, H., Enome, S., {et~al.} 2007, in Coronal Physics from Radio and Space Observations: Proceedings of the CESRA Workshop Held in Nouan le Fuzelier, France, 3--7 June 1996, Springer, 183--191

\bibitem[{Tang {et~al.}(2013)Tang, Wu, \& Tan}]{tang2013electron}
Tang, J., Wu, D., \& Tan, C. 2013, The Astrophysical Journal, 779, 83

\bibitem[{Tingay {et~al.}(2013)Tingay, Goeke, Bowman, Emrich, Ord, Mitchell, Morales, Booler, Crosse, Wayth, {et~al.}}]{tingay2013murchison}
Tingay, S.~J., Goeke, R., Bowman, J.~D., {et~al.} 2013, Publications of the Astronomical Society of Australia, 30, e007

\bibitem[{Trott {et~al.}(2024)Trott, Breen, Green, \& Pearce}]{SKAO-Low-SubStation-Templates}
Trott, C., Breen, S., Green, J., \& Pearce, S. 2024, SKA-Low Substation Templates, doi:10.5281/zenodo.16951143

\bibitem[{{van Haarlem} {et~al.}(2013){van Haarlem}, {Wise}, {Gunst}, {Heald}, {McKean}, {Hessels}, {de Bruyn}, {Nijboer}, {Swinbank}, {Fallows}, {Brentjens}, {Nelles}, {Beck}, {Falcke}, {Fender}, {H{\"o}randel}, {Koopmans}, {Mann}, {Miley}, {R{\"o}ttgering}, {Stappers}, {Wijers}, {Zaroubi}, {van den Akker}, {Alexov}, {Anderson}, {Anderson}, {van Ardenne}, {Arts}, {Asgekar}, {Avruch}, {Batejat}, {B{\"a}hren}, {Bell}, {Bell}, {van Bemmel}, {Bennema}, {Bentum}, {Bernardi}, {Best}, {B{\^\i}rzan}, {Bonafede}, {Boonstra}, {Braun}, {Bregman}, {Breitling}, {van de Brink}, {Broderick}, {Broekema}, {Brouw}, {Br{\"u}ggen}, {Butcher}, {van Cappellen}, {Ciardi}, {Coenen}, {Conway}, {Coolen}, {Corstanje}, {Damstra}, {Davies}, {Deller}, {Dettmar}, {van Diepen}, {Dijkstra}, {Donker}, {Doorduin}, {Dromer}, {Drost}, {van Duin}, {Eisl{\"o}ffel}, {van Enst}, {Ferrari}, {Frieswijk}, {Gankema}, {Garrett}, {de Gasperin}, {Gerbers}, {de Geus}, {Grie{\ss}meier}, {Grit}, {Gruppen}, {Hamaker}, {Hassall}, {Hoeft}, {Holties},
  {Horneffer}, {van der Horst}, {van Houwelingen}, {Huijgen}, {Iacobelli}, {Intema}, {Jackson}, {Jelic}, {de Jong}, {Juette}, {Kant}, {Karastergiou}, {Koers}, {Kollen}, {Kondratiev}, {Kooistra}, {Koopman}, {Koster}, {Kuniyoshi}, {Kramer}, {Kuper}, {Lambropoulos}, {Law}, {van Leeuwen}, {Lemaitre}, {Loose}, {Maat}, {Macario}, {Markoff}, {Masters}, {McFadden}, {McKay-Bukowski}, {Meijering}, {Meulman}, {Mevius}, {Middelberg}, {Millenaar}, {Miller-Jones}, {Mohan}, {Mol}, {Morawietz}, {Morganti}, {Mulcahy}, {Mulder}, {Munk}, {Nieuwenhuis}, {van Nieuwpoort}, {Noordam}, {Norden}, {Noutsos}, {Offringa}, {Olofsson}, {Omar}, {Orr{\'u}}, {Overeem}, {Paas}, {Pandey-Pommier}, {Pandey}, {Pizzo}, {Polatidis}, {Rafferty}, {Rawlings}, {Reich}, {de Reijer}, {Reitsma}, {Renting}, {Riemers}, {Rol}, {Romein}, {Roosjen}, {Ruiter}, {Scaife}, {van der Schaaf}, {Scheers}, {Schellart}, {Schoenmakers}, {Schoonderbeek}, {Serylak}, {Shulevski}, {Sluman}, {Smirnov}, {Sobey}, {Spreeuw}, {Steinmetz}, {Sterks}, {Stiepel}, {Stuurwold},
  {Tagger}, {Tang}, {Tasse}, {Thomas}, {Thoudam}, {Toribio}, {van der Tol}, {Usov}, {van Veelen}, {van der Veen}, {ter Veen}, {Verbiest}, {Vermeulen}, {Vermaas}, {Vocks}, {Vogt}, {de Vos}, {van der Wal}, {van Weeren}, {Weggemans}, {Weltevrede}, {White}, {Wijnholds}, {Wilhelmsson}, {Wucknitz}, {Yatawatta}, {Zarka}, \& {Zensus}}]{lofar_paper}
{van Haarlem}, M.~P., {Wise}, M.~W., {Gunst}, A.~W., {et~al.} 2013, \aap, 556, A2

\bibitem[{White \& Kundu(1997)}]{white1997radio}
White, S., \& Kundu, M. 1997, Solar Physics, 174, 31

\bibitem[{{Wild}(1950)}]{wild}
{Wild}, J.~P. 1950, Australian Journal of Scientific Research A Physical Sciences, 3, 541

\bibitem[{{Wild}(1967)}]{Wild1967}
---. 1967, \pasa, 1, 38

\bibitem[{Winglee \& Dulk(1986)}]{winglee1986electron}
Winglee, R., \& Dulk, G. 1986, Astrophysical Journal, Part 1 (ISSN 0004-637X), vol. 310, Nov. 1, 1986, p. 432-443., 310, 432

\bibitem[{{Yan} {et~al.}(2023){Yan}, {Wu}, {Wu}, {Yang}, {Wu}, {Yan}, \& {Wang}}]{Yan2023-DSRT}
{Yan}, J., {Wu}, J., {Wu}, L., {et~al.} 2023, Nature Astronomy, 7, 750

\bibitem[{Yan {et~al.}(2021)Yan, Chen, Wang, Liu, Geng, Chen, Tan, Chen, Su, \& Tan}]{yan2021mingantu}
Yan, Y., Chen, Z., Wang, W., {et~al.} 2021, frontiers in Astronomy and Space Sciences, 8, 584043

\bibitem[{Zhang {et~al.}(2014)Zhang, Du, Du, \& Sun}]{Zhang2014}
Zhang, Y., Du, A.~M., Du, D., \& Sun, W. 2014, Evaluation of a Revised Interplanetary Shock Prediction Model: 1D CESE-HD-2 Solar-Wind Model, ed. S.~Tomczyk, J.~Zhang, \& T.~Bastian (New York, NY: Springer New York), 537--551

\end{thebibliography}

\end{document}